\documentclass[aps,preprint]{revtex4}%
\usepackage{amsfonts}
\usepackage{amssymb}
\usepackage{graphicx}
\usepackage{color}
\usepackage{amsmath}%
\providecommand{\U}[1]{\protect\rule{.1in}{.1in}}

\begin{document}
\preprint{ }
\title[ ]{Evidence of oblique electron acoustic solitary waves triggered by magnetic reconnection in Earth's magnetosphere}
\author{A. Atteya$^{1,a}$, S. K. EL-Labany$^{2,b}$, P. K. Karmakar$^{3,c}$, M. S. Afify$^{4,d, *}$}
\affiliation{$^{1}$Department of Physics, Faculty of Science, Alexandria University, Alexandria, P.O. 21511, Egypt}
\affiliation{$^{2}$Department of Physics, Faculty of Science, Damietta University, P.O. Box 34517, New Damietta, Egypt}
\affiliation{$^{3}$Department of Physics, Tezpur University, Napaam, Tezpur 784 028, India}
\affiliation{$^{4}$Department of Physics, Faculty of Science, Benha University, Benha 13518, Egypt}

\begin{abstract}
Motivated by the recent Magnetospheric Multiscale (MMS) observations of oblique electron acoustic waves, we addressed the generation mechanism of the observed waves by utilizing the reductive perturbation technique. The nonlinear Zakharov-Kuznetsov (ZK) equation is derived for collisionless, magnetised plasma composed of cool inertial background electrons, the cool inertial electron beam, hot inertialess suprathermal electrons represented by a $\kappa$-distribution, and stationary ions. Moreover, the instability growth rate is derived by using the small-$k$ perturbation expansion method. Our findings reveal that the structure of the electrostatic wave profile is significantly influenced by the external magnetic field, the unperturbed hot, cool, and electron beam densities, the obliquity angle, and the rate of superthermality. Such parameters also affect the instability growth rate. This study clarifies the characteristics of the oblique electron solitary waves that may be responsible for changing the electron and ion distribution functions, which alter the magnetic reconnection process. Moreover, increasing the growth rate with the plasma parameters could be a source of anomalous resistivity that enhances the rate of magnetic reconnection.\\
\textbf{keywords:} Earth's magnetosphere; Oblique electron acoustic solitary waves; Instability growth rate; Zakharov-Kuznetsov equation (ZK); Reductive perturbation technique.

----------------------------------------------------------

$^{\text{*}}$Corresponding author: M. S. Afify

$^{\text{a}}$ahmedatteya@alexu.edu.eg

$^{\text{b}}$skellabany@hotmail.com

$^{\text{c}}$pkk@tezu.ernet.in

$^{\text{d}}$mahmoud.afify@fsc.bu.edu.eg

\end{abstract}
\startpage{1}
\endpage{50}
\date{\today}
\maketitle
\section{Introduction}
\label{Sec. I}
The phenomenon of magnetic reconnection occurs when the frozen-in condition is violated, resulting in the large scale magnetic topology, and the magnetic energy is explosively transformed into kinetic energy that thermalizes the plasma charged particles to high energy. Magnetic reconnection allows plasmas that were previously restricted to regions of different orientations of magnetic fields to mix by changing the magnetic topology \cite{rr1}.
A magnetic reconnection setup can be represented by a separatrix, an X point, an inflow region, an outflow region, and ion and electron diffusion regions. The separatrices include a combination of plasma from both edges of the current sheet and it refers to the newly reconnected magnetic lines. Two separatrices can form an X point. The latter create an X line during three-dimensional (3-D) magnetic reconnection. Ions and electrons are not magnetic in the region of ion and electron diffusion. Moreover, the electron (ion) diffusion region has a width equal to the electron (ion) skin depth ($d_{e, i}=c/\omega_{pe, pi}$), where c, $\omega_{pe}$, and $\omega_{pi}$ are the speed of light, electron plasma frequency, and ion plasma frequency, respectively \cite{r2004, r2020}. 

Addressing the dynamics of magnetic reconnection is important owing to it's impact on the space weather \cite{Froment2021, Lavraud2021} and fusion reactors \cite{Goetz1991}. The energetic particles resulting from the incidence of magnetic reconnection can harm spacecraft equipment and potentially put people at risk, particularly those that are in deep space, where they are not protected by Earth's magnetic field. It's critical to understand how these charged particles are accelerated to such great speeds to predict the short warning time. Further, unexpected sawtooth oscillations and disturbances may result from magnetic reconnection in fusion processes.

Plasma waves are a good candidate to study the characteristics of collisionless magnetic reconnection since they can regulate the particle flux, provide information about the density and composition of the plasma, and carry messages from far plasmas like heliospheric plasma \cite{r1, r2, r3}. Several types of waves have been observed in the regions of magnetic reconnection.
In 2002, a wide frequency spectrum containing electron plasma waves was reported in and near the X-line reconnection region \cite{Farrell2002}. The Geotail satellite and Cluster space craft have also detected additional electrostatic waves in the Earth's magnetotail at the separatrix of the reconnection diffusion area, such as solitay waves, Langumir modes, and electron cyclotron waves. \cite{Deng2004, Viberg2013}. 
Several types of plasma waves have been observed by different space missions such as Geotail and Cluster in the region of Earth's magnetopause associated with magnetic reconnection \cite{Deng2006}. 
Wei et al. \cite{Wei2007} discussed the Cluster mission's observation of waves in the Earth's magnetotail. They demonstrated that the waves are in the range of whistler frequency and that the waves' excitation may be caused by hot electron beams. In order to study the microphysics of magnetic reconnection, the Magnetospheric Multiscale Mission (MMS) was launched in 2015 \cite{Burch2016}. The electron diffusion region at the dayside magnetopause has been successfully encountered by the mission. At the reconnection region in the dayside magnetopause, MMS observed the propagation of different plasma waves such as whistler waves which accompanied by the electrostatic solitary modes \cite{Le Contel2016, Wilder2016, Ergun2016}. MMS has also discovered kinetic Alfven waves and large-amplitude electrostatic waves at the magnetopause, driven on by the mixture of cold and warm plasma close to the electron diffusion area. (\cite{Wilder2019} and references therein). High-frequency electrostatic waves have also been recorded by MMS in the area of the dayside magnetopause's reconnection ion diffusion region \cite{Zhou2016}.

Since waves are believed to introduce anomalous resistivity; i.e. wave particle interaction, in the electron diffusion area, increasing the rate of magnetic reconnection and thermalizing charged particles, the effect plasma waves have on magnetic reconnection is of significant interest to researchers \cite{Khotyaintsev2020}. Drake et al. \cite{Drake2003} documented that electron holes generated by the Buneman instability can cause anomalous resistivity close to the x-line, which can enhance the rate of magnetic reconnection. 
Although recent observations have shown the formation of electron holes near the x-line \cite{Yu2021}, there is no evidence that plasma waves have a direct impact on magnetic reconnection. Furthermore, waves have been proposed to contribute to the onset of magnetic reconnection \cite{Wei2007}, but this has not yet been experimentally confirmed.
Graham et al \cite{Graham2022} utilized the observed parameters and high-resolution fields obtained by MMS to measure cross-field electron diffusion, viscosity, and anomalous resistivity; i.e. wave-particle interaction, caused by lower hybrid waves in the Earth's magnetopause. Their results showed that plasma waves have no contribution to the onset of collisionless magnetic reconnection. However, by making the density gradient steeper, it could impact the rate of magnetic reconnection.
Spacecraft typically go through reconnection regions quickly owing to the fast motion of the reconnecting field lines. The effects of waves on the reconnection process, and in particular the reconnection rate, are consequently exceedingly challenging to experimentally measure.
Therefore, determining the impact of electrostatic waves on magnetic reconnection is challenging. However, we can verify more aspects of the analytical models and simulations using spacecraft data, which will reinforce our trust in their models.

Electrostatic solitary waves are frequently linked to magnetic reconnection at the magnetopause and magnetotail, which can cause the unstable electron distributions needed to form electrostatic solitary waves. Therefore, one of the signs of magnetic reconnection is the presence of electrostatic solitary waves.
For the first time, Cattell et al. \cite{Cattell2002} reported the observation of electrostatic soliatry waves in the Earth's magnetopause. They observed that soliton waves propagated at various rates, equivalent to the fast and slow thermal speeds of electrons and ions, respectively. They argued the difference between soliton modes speeds to the mixing of magnetospheric, magnetosheath, and ionosphere particles that can exist in the magnetopause current layer. 
Reconnection is very asymmetric at the magnetopause of the Earth because the reconnecting plasmas have different densities, temperatures, and magnetic field strengths. As a result, simulations demonstrate that the instabilities that cause electrostatic solitary waves to propagate during asymmetric reconnection can be different from those during symmetric reconnection \cite{Graham2014}. In 2015, Graham et al. \cite{Graham2015} showed that the observed electrostatic waves during the occurrence of magnetic reconnection in the Earth's magnetopause are characterized by distinct speeds in coincidence with Cattell et al. \cite{Cattell2002} observations.
The authors suggested that the Buneman instability; i.e. instability resulting from a relative drift between ions and electrons, bump-on-tail; i.e. instability resulting when an electron beam with a velocity width lower than the beam speed interacts with Maxwellian plasma, and bistreaming instabilities; i.e. instability resulting from two populations of counterstreaming electrons, are reasonable candidate mechanisms for the propagation of electrostatic waves in regions of magnetic reconnection. 

In this work, our effort is to address the mechanism behind the observation of oblique modes in the Earth's magnetopause near the region of magnetic reconnection \cite{Ergun2016}. At present, there is a lack of significant studies of oblique modes at Earth’s magnetopause. As a potential mechanism to produce the oblique modes, we suggested two-stream instability, which results from mixing warm plasma from the magnetosheath with cold magnetospheric plasma, as a possible mechanism to generate the oblique modes. We employed the reductive perturbation technique to derive the Zakharov-Kuznetsov equation (ZK) that can be used to investigate three dimensional problems of oblique electron plasma waves in different plasma arrangements \cite{Zakharov1974}. The solution of ZK provides information about the impact of obliqueness, magnetic field, electron superthermality, etc., on the properties of electrostatic solitary waves.
This work is arranged according to the following way: The physical model and the
derivation of ZK have been presented in sections \ref{Sec. II} and \ref{Sec. III}, respectively. While the solution and the stability of oblique electron soliton waves have been discussed in Sections \ref{Sec. IV} and \ref{Sec. V}, respectively. Then Section \ref{Sec. VI} clarifies the numerical analysis of the current proposed model. Finally, our findings were concluded in Section \ref{Sec. VII}

\section{Basic equations}
\label{Sec. II}

We investigate the generation of electrostatic perturbation in a four-component
collisionless, magnetized plasma of cool inertial electron beam of temperature
$T_{b}$, cool inertial background electrons of temperature $T_{c}$, hot
inertialess suprathermal electrons represented by a $\kappa$-distribution of
temperature $T_{h}$ (the condition $T_{h}\gg T_{b}$ and $T_{c}\neq0$ is satisfied),
and uniformly distributed stationary ions. The external magnetic field $B_{0}$
is considered to be along the z-direction, i.e., $B_{0}=\hat{z}B_{0}$. At
equilibrium, the quasineutrality requirement in such a plasma system is
$n_{i0}=n_{c0}+n_{b0}+n_{h0}$, where $n_{s0}$ is the equilibrium density of
the sth species ($s=h$, $c$, $b$, and $i$ for hot superthermal electrons, cold
electrons, beam electrons, and ions, respectively). The following
normalization equations regulate the fluid model \cite{Danehkar2018}
\begin{equation}
\left.
\begin{array}
[c]{c}%
\frac{\partial n_{c}}{\partial t}+\nabla(n_{c}\mathbf{u}_{c})=0,\\
\frac{\partial n_{b}}{\partial t}+\nabla(n_{b}\mathbf{u}_{b})=0,\\
\frac{\partial\mathbf{u}_{c}}{\partial t}+(\mathbf{u}_{c}.\nabla
)\mathbf{u}_{c}=\nabla\phi-3\theta_{c}n_{c}\nabla n_{c}-\Omega\left(
\mathbf{u}_{c}\times\hat{z}\right)  ,\\
\frac{\partial\mathbf{u}_{b}}{\partial t}+(\mathbf{u}_{b}.\nabla
)\mathbf{u}_{b}=\nabla\phi-3\theta_{b}n_{b}\nabla n_{b}-\Omega\left(
\mathbf{u}_{b}\times\hat{z}\right)  ,\\
\nabla^{2}\phi=\rho^{^{\prime}}=-\left(  1+\rho_{h,c}+\rho_{b,c}\right)
 +n_{c}+\rho_{b,c}n_{b}+\rho_{h,c}\left(
1-\frac{\phi}{\kappa-\frac{3}{2}}\right)  ^{-\kappa+\frac{1}{2}},
\end{array}
\right\}  \tag{1}%
\end{equation}
where $\mathbf{u}_{s}$ is the fluid electron velocity that normalized by the hot
electron thermal velocity $V_{th}=\left(  k_{B}T_{h}/m_{e}\right)  ^{1/2},$ the
potential $\phi$ of the wave which is normalized by $k_{B}T_{h}/e$. The
electron gyro-frequency in normalization form can be $\Omega=\omega_{c}%
/\omega_{pc}$ with $\omega_{c}=eB/m_{e\text{ }}$and $\omega_{pc}=\left(
n_{c0}e^{2}/\epsilon_{0}m_{e}\right)  ^{1/2}$. Also, the quantities
$\rho_{h,c}=n_{h0}/n_{c0}$, $\rho_{b,c}=n_{b0}/n_{c0}$, $\theta_{c}%
=T_{c}/T_{h}$, and $\theta_{b}=T_{b}/T_{h}$. The time variable $t$ is
normalized by $\omega_{pc}^{-1}$ and space variable is normalized by
$\lambda_{0}=\left(  \epsilon_{0}k_{B}T_{h}/n_{c0}e^{2}\right)  ^{1/2}$, with
$k_{B}$ is the Boltzmann constant and $e$ is the electronic charge.
It is critical to note that $kappa > 3/2$ is required for a meaningful particle distribution function. \cite{Hellberg2009}, and the superthermal index $\kappa$
lies in the range $\infty>\kappa>3/2$ \cite{Hellberg2009, Sultana2012, Atteya2018}, where $\kappa
\longrightarrow3/2$ is for extremely superthermal situation and $\kappa
\longrightarrow\infty$ is for Maxwellian case.

\section{Derivation of the ZK equation}
\label{Sec. III}

To obtain the ZK equation, we utilise the reductive perturbation method and
define the stretched independent variables as follows:%
\begin{equation}
X=\epsilon^{\frac{1}{2}}x\text{,\ \ \ \ \ \ \ \ \ \ \ }Y=\epsilon^{\frac{1}%
{2}}y\text{,\ \ \ \ \ \ \ \ \ \ \ \ }Z=\epsilon^{\frac{1}{2}}\left(
L_{z}z-V_{p}t\right)  ,\text{ \ \ \ \ \ \ \ \ \ \ \ \ }\tau=\epsilon^{\frac
{3}{2}}t. \tag{2a}%
\end{equation}
The phase velocity of EAWs is denoted by $V_{p}$. The physical quantities
appearing in Eqs. $\left(  1\right)  $ are extended as power series in
$\epsilon$ about their equilibrium values as:%
\begin{equation}
\left.
\begin{array}
[c]{c}%
n_{c}=\left(  1+\epsilon n_{c}^{\left(  1\right)  }+\epsilon^{2}%
n_{c}^{\left(  2\right)  }+\epsilon^{3}n_{c}^{\left(  3\right)  }+...\right)
,\\
n_{b}=\left(  1+\epsilon n_{b}^{\left(  1\right)  }+\epsilon^{2}%
n_{b}^{\left(  2\right)  }+\epsilon^{3}n_{b}^{\left(  3\right)  }+...\right),
\nonumber\\
u_{cx}=\left(  \epsilon^{\frac{3}{2}}u_{cx}^{\left(  1\right)  }%
+\epsilon^{2}u_{cx}^{\left(  2\right)  }+\epsilon^{\frac{5}{2}}u_{cx}^{\left(
3\right)  }+...\right), \nonumber \\
u_{cy}=\left(  \epsilon^{\frac{3}{2}}u_{cy}^{\left(  1\right)  }%
+\epsilon^{2}u_{cy}^{\left(  2\right)  }+\epsilon^{\frac{5}{2}}u_{cy}^{\left(
3\right)  }+...\right), \nonumber\\
u_{cz}=\left(  \epsilon u_{cz}^{\left(  1\right)  }+\epsilon^{2}%
u_{cz}^{\left(  2\right)  }+\epsilon^{3}u_{cz}^{\left(  3\right)  }+...\right),
\nonumber\\
u_{bx}=\left(  \epsilon^{\frac{3}{2}}u_{bx}^{\left(  1\right)  }%
+\epsilon^{2}u_{bx}^{\left(  2\right)  }+\epsilon^{\frac{5}{2}}u_{bx}^{\left(
3\right)  }+...\right), \nonumber\\
u_{by}=\left(  \epsilon^{\frac{3}{2}}u_{by}^{\left(  1\right)  }%
+\epsilon^{2}u_{by}^{\left(  2\right)  }+\epsilon^{\frac{5}{2}}u_{by}^{\left(
3\right)  }+...\right), \nonumber\\
u_{bz}=\left(  \epsilon u_{bz}^{\left(  1\right)  }+\epsilon^{2}%
u_{bz}^{\left(  2\right)  }+\epsilon^{3}u_{bz}^{\left(  3\right)  }+...\right),
\nonumber\\
\phi=\left(  \epsilon\phi^{\left(  1\right)  }+\epsilon^{2}\phi^{\left(
2\right)  }+\epsilon^{3}\phi^{\left(  3\right)  }+...\right)  .\nonumber
\end{array}
\right\}  \tag{2b}%
\end{equation}
The evolution equations in various order values are discovered by using Eqs.
$(2)$ in Eqs. $(1)$ and gathering distinct powers of $\epsilon$. By solving
the preceding equations, we obtain the following first order relations for
zeroth order in $\epsilon$ as%
\begin{equation}
\left.
\begin{array}
[c]{c}%
n_{c}^{\left(  1\right)  }=\frac{L_{z}^{2}}{3\theta_{c}L_{z}^{2}-V_{p}^{2}%
}\phi^{\left(  1\right)  },\\
n_{b}^{\left(  1\right)  }=\frac{L_{z}^{2}}{3\theta_{b}L_{z}^{2}-V_{p}^{2}%
}\phi^{\left(  1\right)  },\\
u_{cx}^{\left(  1\right)  }=\frac{V_{p}^{2}}{\Omega\left(  3\theta_{c}%
L_{z}^{2}-V_{p}^{2}\right)  }\frac{\partial}{\partial Y}\phi^{\left(
1\right)  },\\
u_{cy}^{\left(  1\right)  }=-\frac{V_{p}^{2}}{\Omega\left(  3\theta_{c}%
L_{z}^{2}-V_{p}^{2}\right)  }\frac{\partial}{\partial X}\phi^{\left(
1\right)  },\\
u_{cz}^{\left(  1\right)  }=\frac{L_{z}V_{p}}{3\theta_{c}L_{z}^{2}-V_{p}^{2}%
}\phi^{\left(  1\right)  },\\
u_{bx}^{\left(  1\right)  }=\frac{V_{p}^{2}}{\Omega\left(  3\theta_{b}%
L_{z}^{2}-V_{p}^{2}\right)  }\frac{\partial}{\partial\zeta}\phi^{\left(
1\right)  },\\
u_{by}^{\left(  1\right)  }=-\frac{V_{p}^{2}}{\Omega\left(  3\theta_{b}%
L_{z}^{2}-V_{p}^{2}\right)  }\frac{\partial}{\partial\zeta}\phi^{\left(
1\right)  },\\
u_{bz}^{\left(  1\right)  }=\frac{L_{z}V_{p}}{3\theta_{b}L_{z}^{2}-V_{p}^{2}%
}\phi^{\left(  1\right)  }.%
\end{array}
\right\}  \tag{3}%
\end{equation}
After plugging Eqs. $(3)$ into Poisson's equation, we get the following
expression for the phase velocity relation:%
\begin{equation}
V_{p}=L_{z}\sqrt{\frac{%
\begin{array}
[c]{c}%
\left(  \left(  2\kappa-3\right)  \left(  1+3\theta_{c}+3\theta_{b}\right)
\left(  1+\rho_{b,c}\right)  +12\rho_{h,c}\left(  \kappa-1\right)  \left(
\theta_{c}+\theta_{b}\right)  \right)  \\
\pm\sqrt{%
\begin{array}
[c]{c}%
-12\left(  \left(  2\kappa-3\right)  \left(  1+\rho_{b,c}\right)  +4\rho
_{h,c}\left(  \kappa-1\right)  \right)  \\
\left(  \left(  2\kappa-3\right)  \left(  \theta_{c}\rho_{b,c}+\theta
_{b}\left(  1+3\theta_{c}\left(  1+\rho_{b,c}\right)  \right)  \right)
+12\left(  \kappa-1\right)  \theta_{c}\theta_{b}\rho_{h,c}\right)  \\
+\left(  \left(  2\kappa-3\right)  \left(  1+3\theta_{c}+3\theta_{b}\right)
\left(  1+\rho_{b,c}\right)  +12\left(  \kappa-1\right)  \left(  \theta
_{c}+\theta_{b}\right)  \rho_{h,c}\right)  ^{2}%
\end{array}
}%
\end{array}
}{2\left(  2\kappa-3\right)  \left(  1+\rho_{h,c}\right)  +8\left(
\kappa-1\right)  \rho_{h,c}}}\tag{4}%
\end{equation}
The presence of two dispersion curves in Eq. $(4)$ indicates that the plasma
model under consideration may transmit cyclotron waves (as fast modes) as well
as acoustic waves (as slow modes). We investigate the influence of important
plasma parameters on slow modes because we are interested in analysing the
characteristics of acoustic modes in the plasma system under consideration.
The phase speed of the slow acoustic wave is dependent on many system factors,
as shown by Eq. $(4)$. We get a collection of second order equations from the
next higher-order of $\epsilon$ and solve them to get%
\begin{equation}
\left.
\begin{array}
[c]{c}%
u_{cx}^{\left(  2\right)  }=\frac{V_{p}^{3}}{\Omega^{2}\left(  3\theta
_{c}L_{z}^{2}-V_{p}^{2}\right)  }\frac{\partial^{2}}{\partial X\partial Z}%
\phi^{\left(  1\right)  },\\
u_{cy}^{\left(  2\right)  }=\frac{V_{p}^{3}}{\Omega\left(  3\theta_{c}%
L_{z}^{2}-V_{p}^{2}\right)  }\frac{\partial^{2}}{\partial Y\partial Z}%
\phi^{\left(  1\right)  },\\
u_{bx}^{\left(  2\right)  }=\frac{V_{p}^{3}}{\Omega^{2}\left(  3\theta
_{b}L_{z}^{2}-V_{p}^{2}\right)  }\frac{\partial^{2}}{\partial X\partial Z}%
\phi^{\left(  1\right)  },\\
u_{by}^{\left(  2\right)  }=\frac{V_{p}^{3}}{\Omega^{2}\left(  3\theta
_{b}L_{z}^{2}-V_{p}^{2}\right)  }\frac{\partial^{2}}{\partial Y\partial Z}%
\phi^{\left(  1\right)  }.%
\end{array}
\right\}  \tag{5}%
\end{equation}
The $\epsilon$ next higher order of equations leads to the following second
order perturbed quantities as%
\begin{equation}
\frac{\partial}{\partial Z}n_{c}^{\left(  2\right)  }=\frac{1}{\Omega
^{2}\left(  3\theta_{c}L_{z}^{2}-V_{p}^{2}\right)  ^{3}}\left(
\begin{array}
[c]{c}%
3\Omega^{2}L_{z}^{4}\left(  L_{z}^{2}\theta_{c}+V_{p}^{2}\right)
\phi^{\left(  1\right)  }\frac{\partial}{\partial Z}\phi^{\left(  1\right)
}\\
+\left(  3\Omega^{2}L_{z}^{4}\left(  2V_{p}^{2}\theta_{c}-3L_{z}^{2}\theta
_{c}^{2}\right)  -L_{z}^{2}\Omega^{2}V_{p}^{4}\right)  \frac{\partial
}{\partial Z}\phi^{\left(  2\right)  }\\
+2L_{z}^{2}\Omega^{2}V_{p}\left(  3\theta_{c}L_{z}^{2}-V_{p}^{2}\right)
\frac{\partial}{\partial T}\phi^{\left(  1\right)  }\\
+V_{p}^{4}\left(  3\theta_{c}L_{z}^{2}-V_{p}^{2}\right)  \left(
\frac{\partial^{3}}{\partial X^{2}\partial Z}\phi^{\left(  1\right)  }%
+\frac{\partial^{3}}{\partial Y^{2}\partial Z}\phi^{\left(  1\right)
}\right)
\end{array}
\right)  ,\tag{6a}%
\end{equation}

\begin{equation}
\frac{\partial}{\partial Z}n_{b}^{\left(  2\right)  }=\frac{1}{\Omega
^{2}\left(  3\theta_{b}L_{z}^{2}-V_{p}^{2}\right)  ^{3}}\left(
\begin{array}
[c]{c}%
3\Omega^{2}L_{z}^{4}\left(  L_{z}^{2}\theta_{b}+V_{p}^{2}\right)
\phi^{\left(  1\right)  }\frac{\partial}{\partial Z}\phi^{\left(  1\right)
}\\
+\left(  3\Omega^{2}L_{z}^{4}\left(  2V_{p}^{2}\theta_{b}-3L_{z}^{2}\theta
_{b}^{2}\right)  -L_{z}^{2}\Omega^{2}V_{p}^{4}\right)  \frac{\partial
}{\partial Z}\phi^{\left(  2\right)  }\\
+2L_{z}^{2}\Omega^{2}V_{p}\left(  3\theta_{b}L_{z}^{2}-V_{p}^{2}\right)
\frac{\partial}{\partial T}\phi^{\left(  1\right)  }\\
+V_{p}^{4}\left(  3\theta_{b}L_{z}^{2}-V_{p}^{2}\right)  \left(
\frac{\partial^{3}}{\partial X^{2}\partial Z}\phi^{\left(  1\right)  }%
+\frac{\partial^{3}}{\partial Y^{2}\partial Z}\phi^{\left(  1\right)
}\right)
\end{array}
\right)  ,\tag{6b}%
\end{equation}
The ZK equation may be built by equating these variables in Poisson's equation
and making certain algebraic adjustments to yield the following form%
\begin{equation}
\frac{\partial\phi_{1}}{\partial\tau}+A\phi_{1}\frac{\partial\phi_{1}%
}{\partial Z}+B\frac{\partial^{3}\phi_{1}}{\partial Z^{3}}+C\frac{\partial
}{\partial Z}\left(  \frac{\partial^{2}}{\partial X^{2}}+\frac{\partial^{2}%
}{\partial Y^{2}}\right)  \phi_{1}=0, \tag{7}%
\end{equation}
where the nonlinear coefficient (A), the dispersion coefficient (B), and the coefficient (C) are given as:
\begin{align}
A  & =\frac{\left(  3\theta_{c}L_{z}^{2}-V_{p}^{2}\right)  ^{2}\left(
3\theta_{b}L_{z}^{2}-V_{p}^{2}\right)  ^{2}}{2L_{z}^{2}V_{p}\left(
3\theta_{b}L_{z}^{2}-V_{p}^{2}\right)  ^{2}+\rho_{b,c}\left(  3\theta_{c}%
L_{z}^{2}-V_{p}^{2}\right)  ^{2}}\times\nonumber\\
& \left(  1+\rho_{b,c}+3L_{z}^{4}\left(  \frac{V_{p}^{2}+\theta_{c}L_{z}^{2}%
}{\left(  3\theta_{c}L_{z}^{2}-V_{p}^{2}\right)  ^{3}}+\frac{\left(  V_{p}%
^{2}+\theta_{b}L_{z}^{2}\right)  \rho_{b,c}}{\left(  3\theta_{b}L_{z}%
^{2}-V_{p}^{2}\right)  ^{3}}\right)  +\frac{2\left(  5-6\kappa\right)
}{\left(  3-2\kappa\right)  ^{2}}\rho_{h,c}\right)  \tag{8a}%
\end{align}%
\begin{equation}
B=\frac{\left(  3\theta_{c}L_{z}^{2}-V_{p}^{2}\right)  ^{2}\left(  3\theta
_{b}L_{z}^{2}-V_{p}^{2}\right)  ^{2}}{2L_{z}^{2}V_{p}\left(  3\theta_{b}%
L_{z}^{2}-V_{p}^{2}\right)  ^{2}+\rho_{b,c}\left(  3\theta_{c}L_{z}^{2}%
-V_{p}^{2}\right)  ^{2}}\tag{8b}%
\end{equation}%
\begin{equation}
C=\frac{\left(  3\theta_{c}L_{z}^{2}-V_{p}^{2}\right)  ^{2}\left(  3\theta
_{b}L_{z}^{2}-V_{p}^{2}\right)  ^{2}}{2L_{z}^{2}V_{p}\left(  3\theta_{b}%
L_{z}^{2}-V_{p}^{2}\right)  ^{2}+\rho_{b,c}\left(  3\theta_{c}L_{z}^{2}%
-V_{p}^{2}\right)  ^{2}}\left(  1+\frac{V_{p}^{4}}{\Omega^{2}}\left(  \frac
{1}{\left(  3\theta_{c}L_{z}^{2}-V_{p}^{2}\right)  ^{2}}+\frac{\rho_{b,c}%
}{\left(  3\theta_{b}L_{z}^{2}-V_{p}^{2}\right)  ^{2}}\right)  \right)
\tag{8c}%
\end{equation}

\section{Solitary wave analysis}
\label{Sec. IV}

The solitary wave solution of Eq. $(7)$ results from the balance between
nonlinearity and dispersive effects, we will first investigate the
transformation of the independent variables \cite{Allen1993, Allen1995, Mamun1998} by
rotating the spatial axes ($X$, $Z$) by an angle $\theta$ and renaming $Y$ and
$T$, the coordinate transformations are described as follows:
\begin{equation}
\left.
\begin{array}
[c]{c}%
\zeta=X\cos\theta-Z\sin\theta,\\
\xi=X\sin\theta+Z\cos\theta,\\
\eta=Y,\text{and }\tau=T.
\end{array}
\right\}  \tag{9}%
\end{equation}
When we apply these changes to the ZK Eq. $(7)$, we get%
\begin{equation}
\left.
\begin{array}
[c]{c}%
\frac{\partial\phi^{(1)}}{\partial\tau}+R_{1}\phi^{(1)}\frac{\partial
\phi^{(1)}}{\partial\xi}+R_{2}\frac{\partial^{3}\phi^{(1)}}{\partial\xi^{3}%
}+R_{3}\phi^{(1)}\frac{\partial\phi^{(1)}}{\partial\zeta}+R_{4}\frac
{\partial^{3}\phi^{(1)}}{\partial\zeta^{3}}\\
+R_{5}\frac{\partial^{3}\phi^{(1)}}{\partial\xi^{2}\partial\zeta}+R_{6}%
\frac{\partial^{3}\phi^{(1)}}{\partial\xi\partial\zeta^{2}}+R_{7}%
\frac{\partial^{3}\phi^{(1)}}{\partial\xi\partial\eta^{2}}+R_{8}\frac
{\partial^{3}\phi^{(1)}}{\partial\zeta\partial\eta^{2}}=0,
\end{array}
\right\}  \tag{10}%
\end{equation}
where the coefficients in Eq. (10) are given as:
\begin{equation}
\left.
\begin{array}
[c]{c}%
R_{1}=A\cos\theta,R_{2}=B\cos^{3}\theta+C\sin^{2}\theta\cos\theta,\\
R_{3}=-A\sin\theta,R_{4}=-B\sin^{3}\theta-C\cos^{2}\theta\sin\theta,\\
R_{5}=2C(\sin\theta\cos^{2}\theta-\frac{1}{2}\sin^{3}\theta)-3B\cos^{2}%
\theta\sin\theta,\\
R_{6}=-2C(\sin^{2}\theta\cos\theta-\frac{1}{2}\cos^{3}\theta)+3B\sin^{2}%
\theta\cos\theta,\\
R_{7}=C\cos\theta,R_{8}=-C\sin\theta.
\end{array}
\right\}  \tag{11}%
\end{equation}
The steady-state solution of the ZK equation is as follows:%
\[
\phi^{(1)}=\phi_{0}(\rho),
\]
where $\rho=\xi-M\tau$, and $M$ is the Mach number normalized by the
ion-acoustic speed $c_{i}$. So, Eq. $\left(  10\right)  $\ can be
expressed as \cite{Haider2012}
\begin{equation}
-M\frac{d\phi_{0}}{d\rho}+R_{1}\phi_{0}\frac{d\phi_{0}}{d\rho}+R_{2}%
\frac{d^{3}\phi_{0}}{d\rho^{3}}=0. \tag{12}%
\end{equation}
We derive the EASWs pulse solution by integrating and applying suitable
boundary conditions
\begin{equation}
\phi_{0}(\rho)=\phi_{m}\operatorname{sech}^{2}(\frac{\rho}{w}), \tag{13}%
\end{equation}
where $\varphi_{m}$ and $w$ are the solitary wave amplitude and width,
respectively; these can be shown as
\begin{equation}
\phi_{m}=3M/R_{1}\text{ and }w=2\sqrt{R_{2}/M}. \tag{14}%
\end{equation}
The accompanying electric field might be deducted and have the following form:
\begin{equation}
E_{0}(\rho)=-\bigtriangledown\phi_{0}(\rho)=\frac{2\phi_{m}}{w}%
\operatorname{sech}^{2}(\frac{\rho}{w})\tanh(\frac{\rho}{w}). \tag{15}%
\end{equation}
It is clear from Eqs. $(11)$ and $(14)$ that both the amplitude and the width
of the solitary wave depend on the cold, and beam electrons densities and
temperatures. Also, the superthermality and obliquness have considerable
effects. The wave amplitude is an essential element to describe the energy of the wave according to the following equation \cite{Afify2021, El-Monier2021}:
\begin{align}
E_{n}  &  =%
{\displaystyle\int\limits_{-\infty}^{\infty}}
\frac{\phi_{0}^{2}(\rho)}{\lambda^{2}}d\rho,\nonumber\\
E_{n}  &  =\frac{4w\phi_{m}^{2}}{3\lambda^{2}}. \tag{16}%
\end{align}
It displays the amount to which the confined charged particles obtain energy
from the wave.

\section{Instability Analysis}
\label{Sec. V}

We use the small-k expansion perturbation approach to investigate the
stability of this solution. We assume the following expression for the perturbed electric potential \cite{Mamun1998, El-Labany2013}:
\begin{equation}
\phi^{\left(  1\right)  }=\phi_{0}(\rho)+\Phi(\rho,\zeta,\eta,\tau), \tag{17}%
\end{equation}
where the inclined plane spreading long-wavelength-wave takes the symbol
$\Phi$ that can be represented as%
\begin{equation}
\Phi(\rho,\zeta,\eta,\tau)=\psi(\rho)\exp i[k(l_{\xi}\rho+l_{\zeta}%
\zeta+l_{\eta}\eta)-\gamma\tau], \tag{18}%
\end{equation}
in which $l_{\xi}^{2}+l_{\zeta}^{2}+l_{\eta}^{2}=1$, $\psi(\rho)$, and
$\gamma$ can be enhanced by applying small values of $k$ to the form%
\begin{equation}
\left.
\begin{array}
[c]{c}%
\psi(\rho)=\psi_{o}+k\psi_{1}+k^{2}\psi_{2}+...,\\
\gamma=k\gamma_{1}+k^{2}\gamma_{2}+....
\end{array}
\right\}  \tag{19}%
\end{equation}
The linearized ZK equation can be obtained by substituting Eq. $(17)$ into Eq.
$(10)$ to be as%
\begin{equation}
\left.
\begin{array}
[c]{c}%
\frac{\partial\Phi}{\partial\tau}-M\frac{\partial\Phi}{\partial\rho}+R_{1}%
\phi_{0}\frac{\partial\Phi}{\partial\rho}+R_{2}\frac{\partial^{3}\Phi
}{\partial\rho^{3}}\\
+R_{3}\phi_{0}\frac{\partial\Phi}{\partial\zeta}+R_{4}\frac{\partial^{3}\Phi
}{\partial\zeta^{3}}+R_{5}\frac{\partial^{3}\Phi}{\partial\rho^{2}%
\partial\zeta}+R_{6}\frac{\partial^{3}\Phi}{\partial\rho\partial\zeta^{2}%
}+R_{7}\frac{\partial^{3}\Phi}{\partial\rho\partial\eta^{2}}+R_{8}%
\frac{\partial^{3}\Phi}{\partial\zeta\partial\eta^{2}}=0.
\end{array}
\right\}  \tag{20}%
\end{equation}
Substituting Eqs. $(18)$ and $(19)$ into Eq. $(20)$ and equating the
same-power coefficients of the $k$ order,  resulting in
\begin{equation}
(-M+R_{1}\phi_{0})\psi_{o}+R_{2}\frac{d^{2}\psi_{o}}{d\rho^{2}}=C^{^{\prime}},
\tag{21}%
\end{equation}
$C^{^{\prime}}$ denotes the integration constant. The homogeneous part of Eq.
($21$) has two independent linear solutions which can be expressed as \cite{Mamun1998}:
\begin{equation}
f=\frac{d\psi_{0}}{d\rho},g=f\int^{\rho}\frac{d\rho}{f^{2}}. \tag{22}%
\end{equation}
As a result, the generic solution might take the following form
\begin{equation}
\psi_{0}=C_{1}f+C_{2}g-C^{^{\prime}}f\int^{\rho}\frac{g}{S_{2}}d\rho
+C^{^{\prime}}g\int^{\rho}\frac{f}{S_{2}}d\rho, \tag{23}%
\end{equation}
where $C_{1}$ and $C_{2}$ are the constants of the integration. Finally, the zeroth-order general solution equation may be reduced to the
following equation%
\begin{equation}
\psi_{0}=C_{1}f. \tag{24}%
\end{equation}
The first and second-order equations may be found from Eqs. $(18)-(20)$, and
the dispersion relation can be represented by their solutions as follows%
\begin{equation}
\gamma_{1}=\Delta-Ml_{\xi}+\sqrt{\Delta^{2}-\Gamma}, \tag{25}%
\end{equation}
where%
\begin{equation}
\left.
\begin{array}
[c]{c}%
\Delta=\frac{2}{3}(\mu_{1}\phi_{m}-2\mu_{2}/w^{2}),\\
\Gamma=\frac{16}{45}(\mu_{1}^{2}\phi_{m}^{2}-3\mu_{1}\mu_{2}\phi_{m}%
/w^{2}-3\mu_{2}^{2}/W^{4}+12R_{2}\mu_{3}/w^{4}),\\
\mu_{1}=(R_{1}l_{\xi}+R_{3}l_{\zeta})\text{, }\mu_{2}=(3R_{2}l_{\xi}%
+R_{5}l_{\zeta}),\\
\text{and }\mu_{3}=(3R_{2}l_{\xi}^{2}+2R_{5}l_{\xi}l_{\zeta}+R_{6}l_{\zeta
}^{2}+R_{7}l_{\eta}^{2}).
\end{array}
\right\}  \tag{26}%
\end{equation}
As a result of Eq. $(25)$, we can see that instability arises when the
condition $\Gamma-\Delta^{2}>0$ is met. We calculate the instability growth
rate, $Gr$, which is denoted as%
\begin{equation}
Gr=\sqrt{\Gamma-\Delta^{2}}. \tag{27}%
\end{equation}
The impact of the plasma parameters on the growth rate will be discussed in the next section.

\section{Numerical analysis and discussion}
\label{Sec. VI}

In this section, we employed the plasma parameters observed by the MMS mission in the Earth's magnetopause, to address the generation mechanism and instability of oblique electron plasma waves close to the magnetic reconnection domain \cite{Ergun2016, Graham2016}.
Using the reductive perturbation technique, we deduced the dispersion relation then the nonlinear analysis yields the 3+1 dimension ZK equation in a four-component collisionless, magnetised plasma composed of cold inertial background electrons, cool inertial electron beam, hot inertialess suprathermal electrons, and evenly distributed stationary ions. The solution to this developed equation yields a solitary wave, then we obtain its associated electric field and energy. The small-$k$ expansion approach was used to investigate the multidimensional instability of the earlier ZK equation \cite{Fortov2005, Zedan2020}.

The results of this investigation may be summarized as follows:
The phase velocity, $V_{p}$, of the considered plasma model indicates the
presence of two modes. These ensure access to acoustic waves and
cyclotron waves to propagate in slow and fast modes, respectively. This phase velocity is affected by the ratio of hot to cool electron number density, $\rho_{h,c}$, the ratio of beam to cool electron number density, $\rho_{b,c}$, the ratio of cool to hot
electron temperature, $\theta_{c}$, the ratio of beam to hot electron
temperature, $\theta_{b}$, the obliquity angle, $\theta$, and the
superthermal parameter, $\kappa$.
Figure (1a) shows the variation of phase velocity with the obliquity angle at different values of superthermality. It is clear that the phase velocity of EASWs decreases as the obliquity angle is increased. However, increasing the value of superthermality leads to an increase in the phase velocity, following the same behaviour as with the variation of the obliquity angle.
The effect of the beam to hot electron temperature ratio is dominant at its small values while its increase leads to the propagation of oblique ESWs with distinguishable velocities at certain values of the cold electron temperature ratio as depicted in Fig. (1b). This means that at a certain value of the electron beam temperature, the instability grows, giving rise to the ability to move the energy from the electron beam to the propagated electrostatic waves.
We can observe from Fig. (1c) that the phase velocity has a slight decrease with the increase of the beam to cold electron number density ratio. Investigating this result at different values of the hot to cold electron number density ratio showed a significant reduction in the wave phase velocity. This means that different instabilities are responsible for the generation of oblique ESWs. This is because at very small values of the hot to cold electron number density, the two stream instability and the Buneman instabilities are a reasonable source of the electrostatic waves. While increasing the value of hot to cold electron number density with respect to the beam to electron number density ratio causes the bump-on-tail instability to be domainant. Therefore, the oblique ESWs could propagate with different velocities near the regions of magnetic reconnection depending on the type of instability in accordance with results by Graham et al. \cite{Graham2016}.
The nonlinear term (A) controls the steepening and polarity of the generated
waves. Figure (2) depicts the dependency of the nonlinear coefficient, (A), on
the previously indicated parameters. It is shown that (A) can be either positive
or negative depending on the plasma parameters. The positive (A) is related with compressive EASWs, whereas the negative (A) is associated with rarefactive EASWs. 
We discovered that increasing both the obliquity angle and beam to cold electron number density lead to a decrease in the coefficient (A) at different values of superthermality and hot to cold number density ratio according to Figs. (2a) and (2c), respectively. As illustrated in Fig. (2b), for high levels of beam to hot electron temperature ratio, the coefficient (A) becomes negative, and its absolute value grows as the cold electron temperature increases.
The effects of $\kappa$, $\theta_{c}$, and $\rho_{h,c}$ on the properties of longitudinal dispersion coefficient (B) during the variation with $\theta$, $\theta_{b}$, and $\rho_{b,c}$ are manifested in Figs. (3a), (3b), and (3c), respectively. We noticed that the dispersion coefficient (B) decreases as the obliquity angle increases, while it increases as ( $\theta_{b}$ and $\rho_{b,c}$) increase. 
Figures (4) and (5) present the variation of solitary wave amplitude $(\phi_{m})$ and width $(w)$ with the prior factors, respectively. Such electrostatic waves develop when the nonlinear and dispersion coefficients are balanced. Since the amplitude depends on the nonlinear term, (A), the amplitude can be positive or negative depending on the values of the system parameters. For smaller values of the cool to hot electron temperature ratio and the hot to cool electron number density, the amplitude becomes smaller as presented by Figs. (4b) and (4c). Moreover, the pulse width (w) is suppressed by increasing the values of the hot to cold electron number density ratio according to Fig. (5c).
The results obtained from Figs. (4) and (5) are confirmed by the solitary wave profiles and the associated electric field as shown in Figs. (6) and (7), respectively. Further, we can observe from Fig. (6d) that increasing the electron gyro-frequency leads to an increase in the pulse width without a significant change in the pulse amplitude.
Figure (8) shows the variation of wave energy, which is mostly determined by the wave amplitude with the same parameters impact $(\phi_{m})$. It is clear that energy is controlled by the same parameters. As a result, the energy incorporates the impacts of the preceding parameters on $\phi_{m}$.
The variation of the instability growth rate (Gr) against $\theta$, $\theta_{c}$, and $\rho_{b,c}$ is illustrated in Fig. (9). It is obvious from Fig. (9a) that (Gr) decreases as $\theta$ increases, and it reaches zero at a certain value of $\theta$. This critical value depends on the superthermal parameter. Figure (9b) shows that the reduction of $Gr$ becomes sharp after $\theta_{c}$ attains a certain value. This critical value of $\theta_{c}$ depends on $\theta_{b}$.
Moreover, Fig. (9c) demonstrates that the growth rate (Gr) increases with increasing the beam to cold electron number density at a certain value of the hot to cool electron number density. However, increasing the value of $\rho_{h,c}$ leads to reduce the growth rate. Therefore, regions with excess cold electrons could be a source of anomalous resistivity that could affect the activation of magnetic reconnection (\cite{PARIS1973, Shukla2002, Shi2020, Mayur2022} and references therein). 

\section{conclusions}
\label{Sec. VII}

In this work, we employed the multifluid model to address the characteristics of oblique electron acoustic solitary waves observed by the MMS mission in Earth's magnetopause near the regions of magnetic reconnection. Our efforts are also to investigate the connection between electrostatic waves and the rate of magnetic reconnection.
We derived the ZK equation to explain nonlinear small-amplitude electron-acoustic waves in a collisionless, magnetized plasma with four components: cold inertial background electrons; a cool inertial electron beam; hot inertialess suprathermal electrons; and evenly distributed stationary ions. The solution of the ZK equation has been used to investigate the properties of EASWs. Our results may be summed up as follows:
\begin{itemize}
\item This study provides the possibility of the generation of both positive and negative oblique electron acoustic waves in the Earth's magnetopause. 
\item We observed that the plasma parameters such as the obliquity angle, the ratio of beam to hot electron temperature, the ratio of cool to hot electron temperature, the ratio of beam to cold electron number density, and the ratio of hot to cold electron number density had a significant effect on the structure of the soliton profile. 
\item We discovered that the oblique soliton waves could propagate with different velocities close to regions of magnetic reconnection. The Buneman instability and the two-stream instability, which are related to the properties of the electron beam velocity, and the bump-on-tail instability, which is related to the density of the background plasma, are responsible for the difference in the phase velocity of the oblique ESWs.
\item The influence of plasma parameters on the instability growth rate has been explored. It has been demonstrated that increasing the ratio of cold to hot electron density and obliqueness can slow the rate of instability progression. While increasing the ratio of beam to cold electron number density leads to an increase in the instability development rate. 
\end{itemize}
Finally, we stressed that regions with excess cold electrons can be a source of anomalous resistivity, which might have an impact on the rate of magnetic reconnection. The role of electrostatic solitary waves in generating whistler waves in regions of magnetic reconnection and the microphysics of the connection between electrostatic soliatry waves and the activation of magnetic reconnection will be addressed in the future. 


\newpage

\begin{figure}[!tbp]
  \centering
  \begin{minipage}[b]{0.45\textwidth}
    \includegraphics[width=\textwidth]{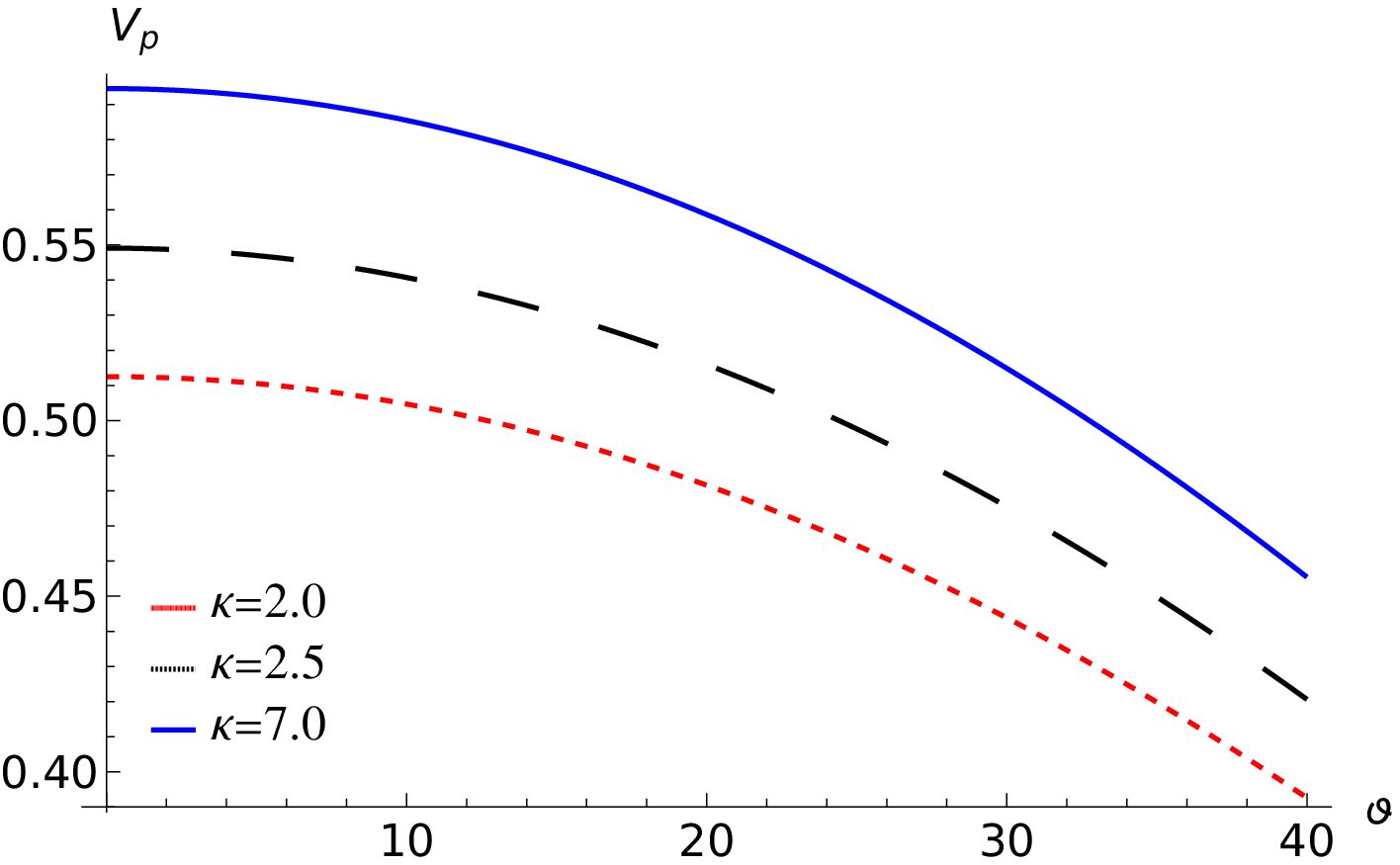}
  \end{minipage}
  \hfill
  \begin{minipage}[b]{0.45\textwidth}
    \includegraphics[width=\textwidth]{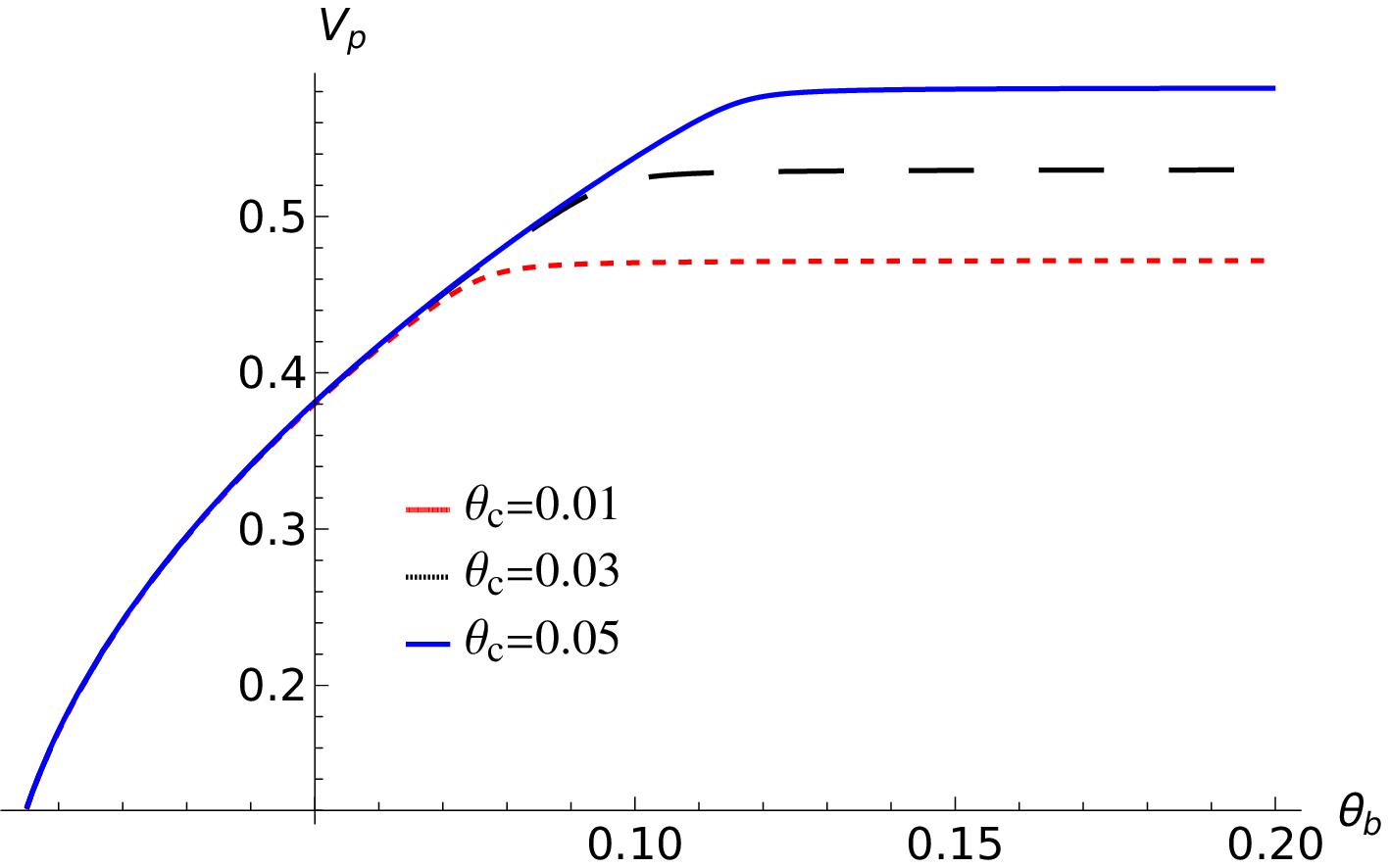}
  \end{minipage}
    \centering
	\includegraphics[width=0.6\linewidth]{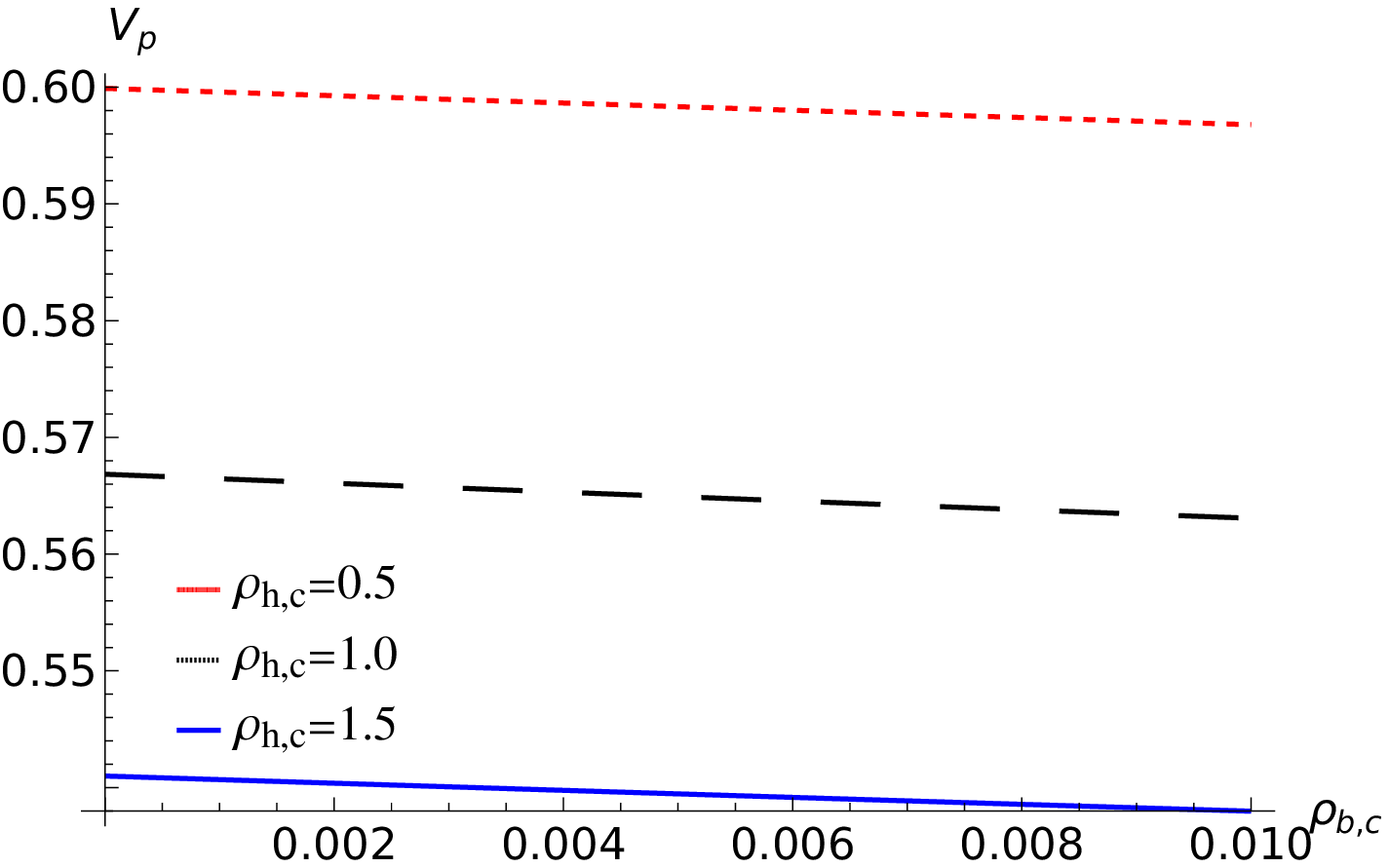}
	\caption{The variation of the phase velocity
$V_{p}$, represented by Eq. $\left(  4\right)  $ a) versus $\vartheta$ for
different values of $\kappa$ at $\rho_{h,c}=1.5$, $\rho_{b,c}=0.003$,
$\theta_{c}=0.04$ with $\theta_{b}=0.25$, b) against $\theta_{b}$ for
different values of $\theta_{c}$ at $\rho_{h,c}=1.5$, $\rho_{b,c}=0.003$,
$\kappa=3$ with $\vartheta=5.0$, c) against $\rho_{b,c}$ for different values
of $\rho_{h,c}$ at $\kappa=3.0$, $\vartheta=5.0$, $\theta_{c}=0.03$ with
$\theta_{b}=0.15$.}
\label{Figure1}
\end{figure}
\begin{figure}[!tbp]
  \centering
  \begin{minipage}[b]{0.45\textwidth}
    \includegraphics[width=\textwidth]{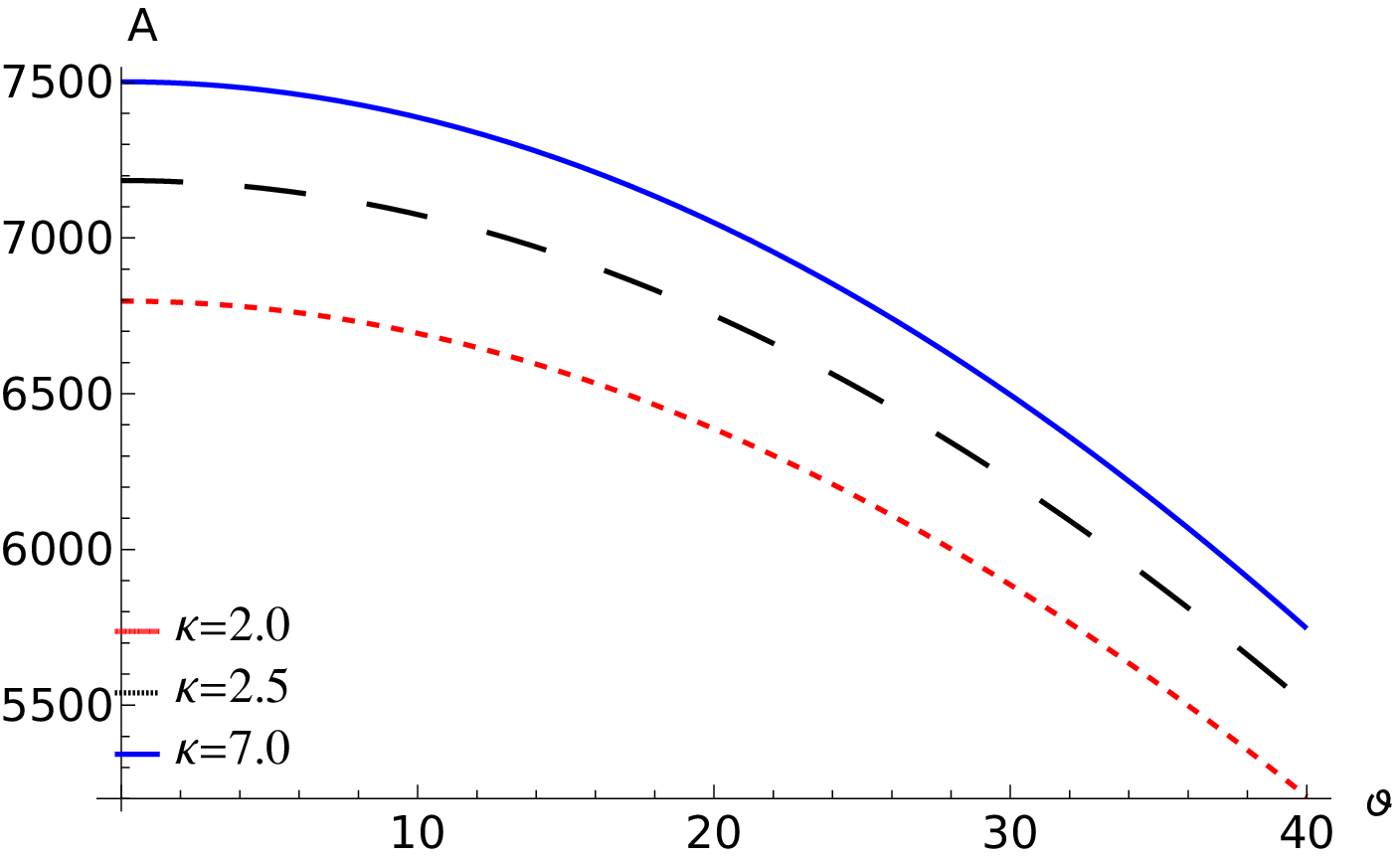}
  \end{minipage}
  \hfill
  \begin{minipage}[b]{0.45\textwidth}
    \includegraphics[width=\textwidth]{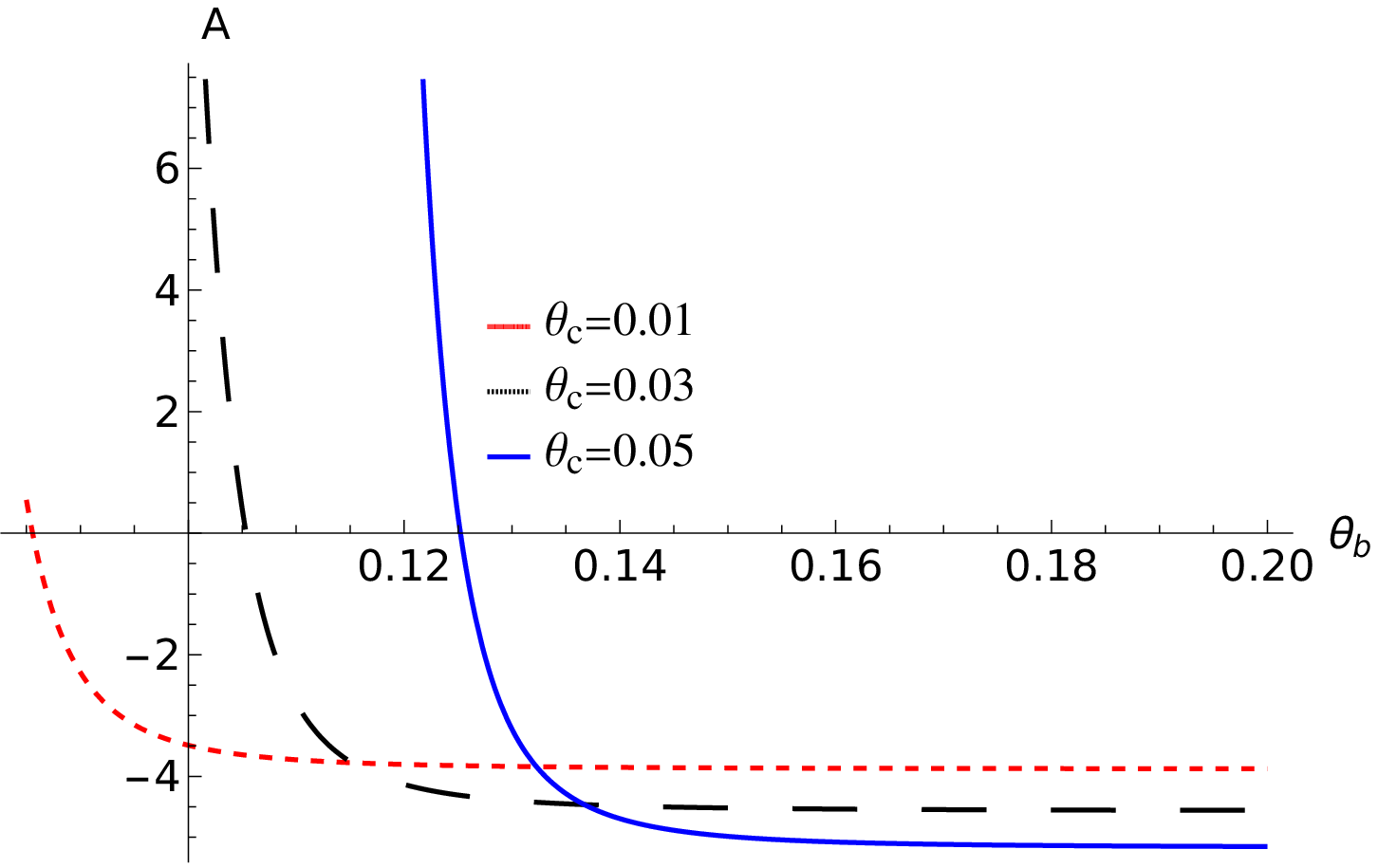}
  \end{minipage}
    \centering
	\includegraphics[width=0.6\linewidth]{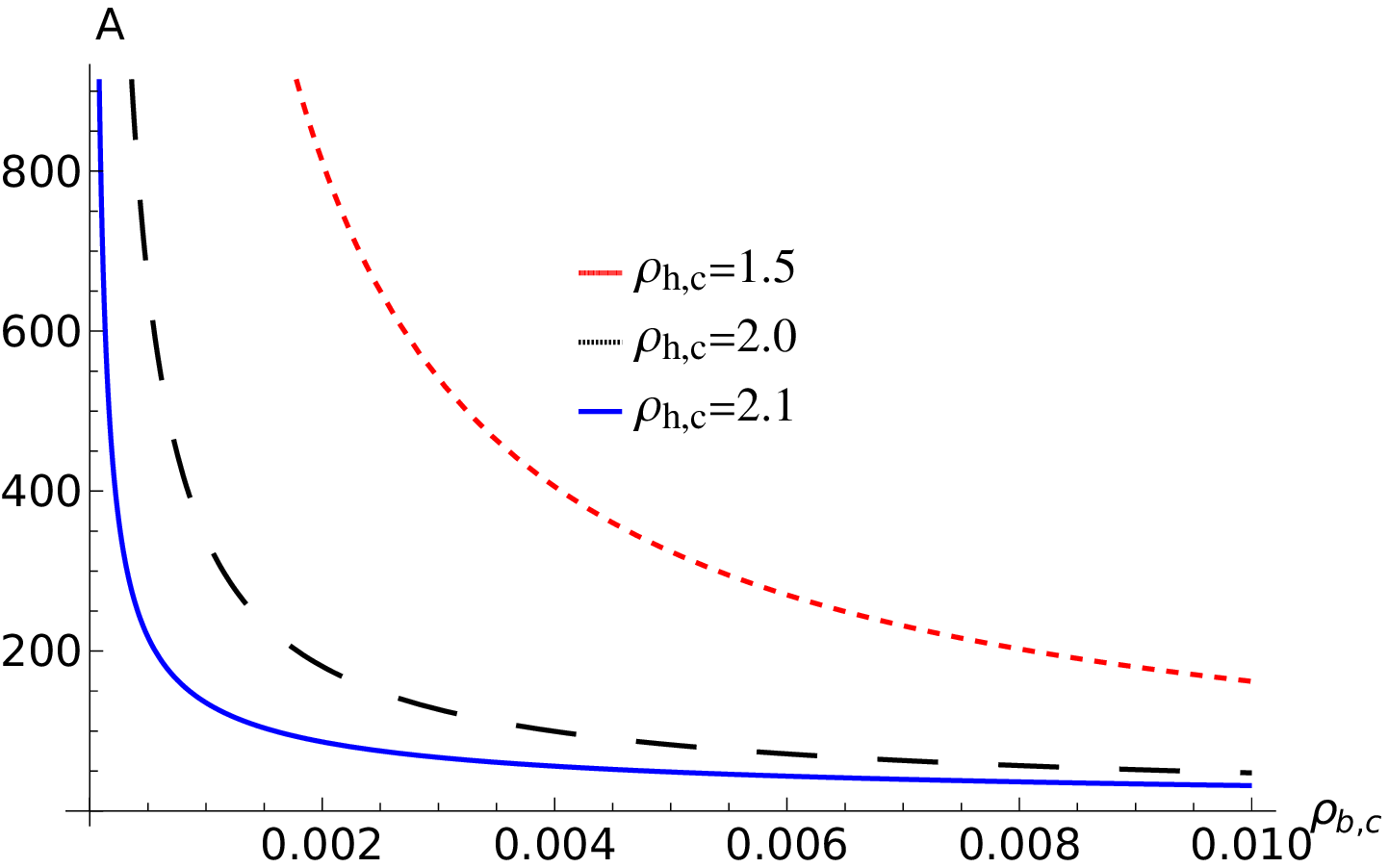}
	\caption{The variation of the nonlinear term
$A$, represented by equation $\left(  8a\right)  $ a) against $\vartheta$ for
different values of $\kappa$ at $\rho_{h,c}=1.5$, $\rho_{b,c}=0.003$,
$\theta_{c}=0.04$ with $\theta_{b}=0.05$, b) against $\theta_{b}$ for
different values of $\theta_{c}$ at $\rho_{h,c}=1.5$, $\rho_{b,c}=0.003$,
$\kappa=3$ with $\vartheta=5.0$, c) against $\rho_{b,c}$ for different values
of $\rho_{h,c}$ at $\kappa=3.0$, $\vartheta=5.0$, $\theta_{c}=0.03$ with
$\theta_{b}=0.08$.}
\label{Figure2}
\end{figure}
\begin{figure}[!tbp]
  \centering
  \begin{minipage}[b]{0.45\textwidth}
    \includegraphics[width=\textwidth]{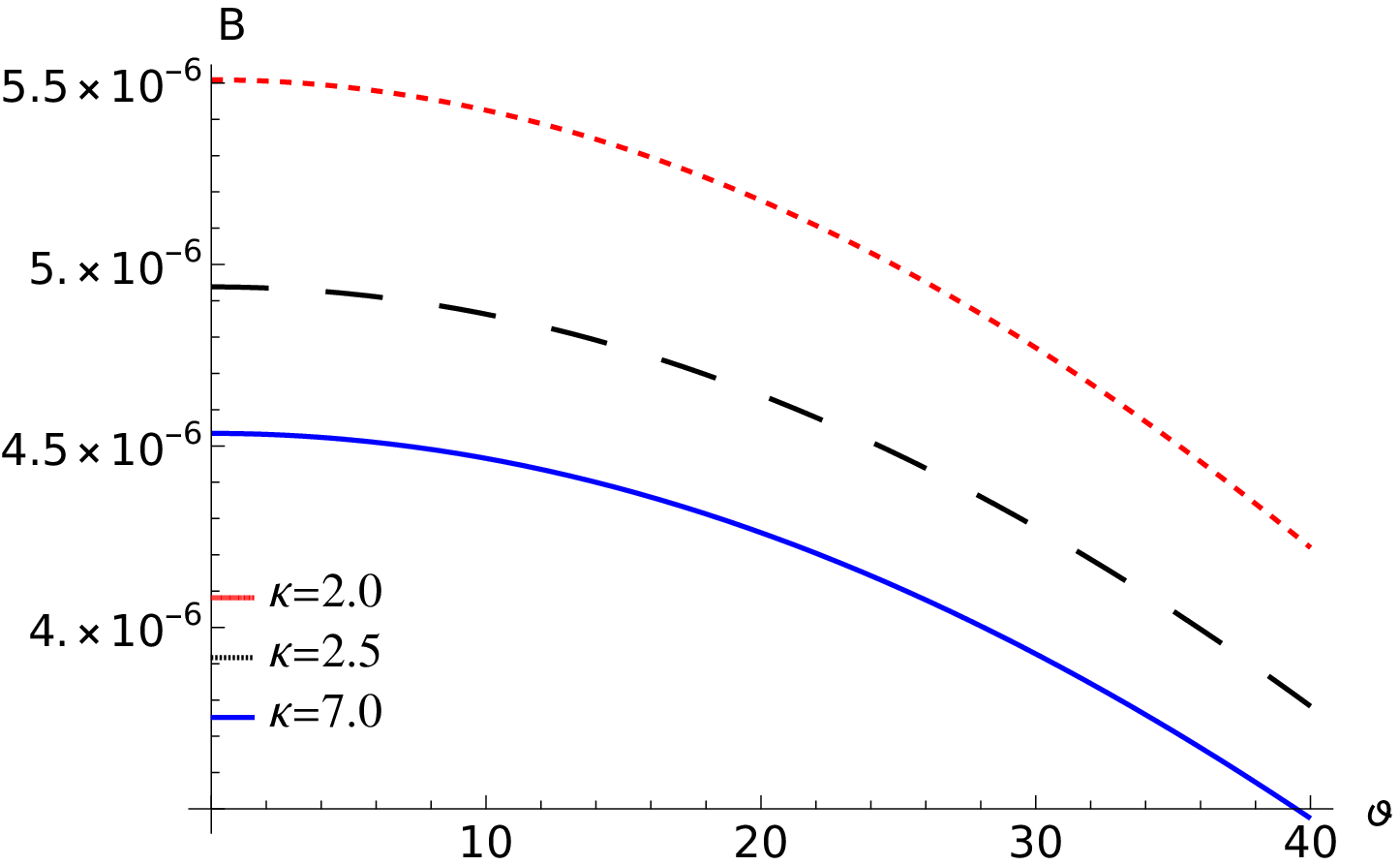}
  \end{minipage}
  \hfill
  \begin{minipage}[b]{0.45\textwidth}
    \includegraphics[width=\textwidth]{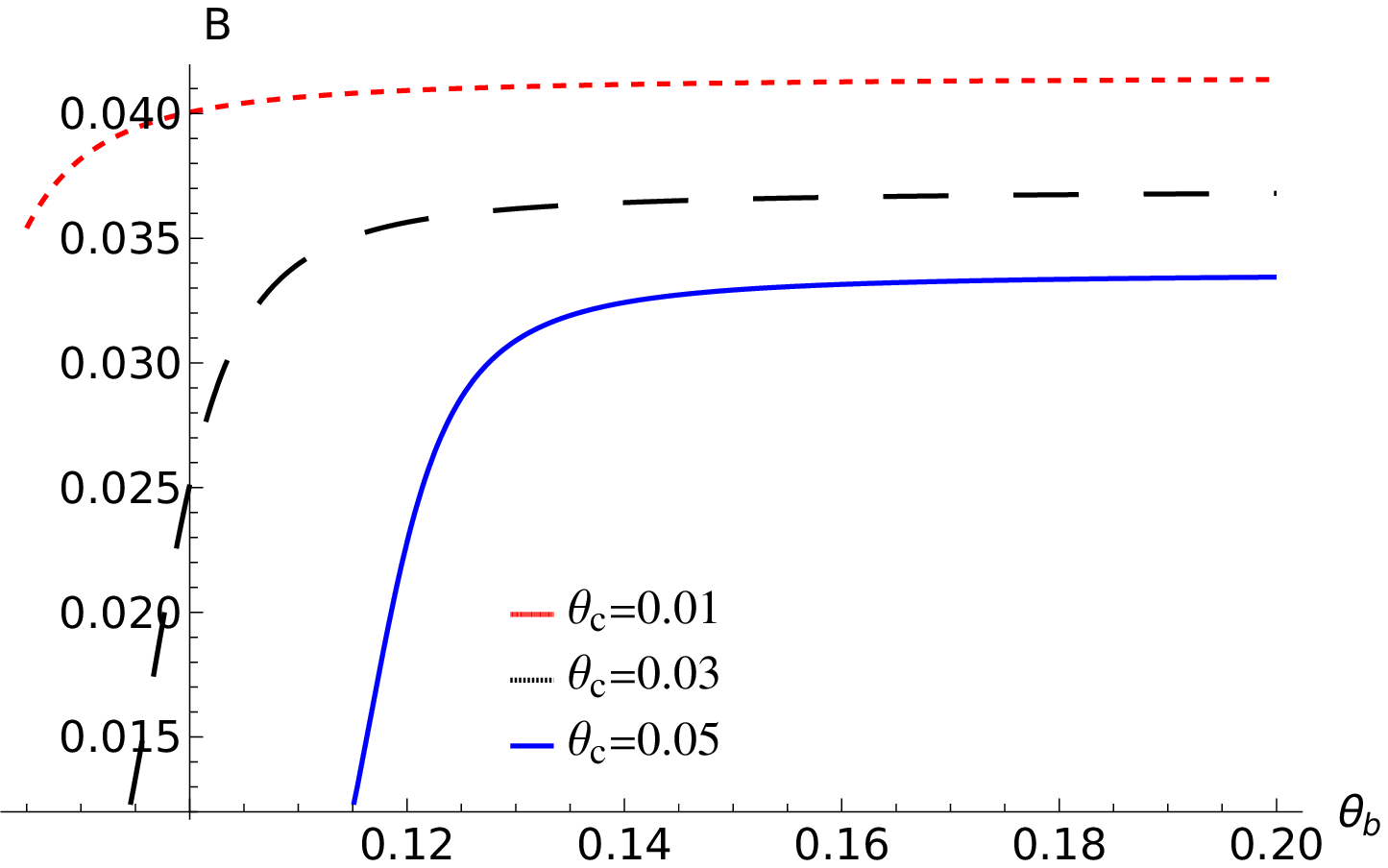}
  \end{minipage}
    \centering
	\includegraphics[width=0.6\linewidth]{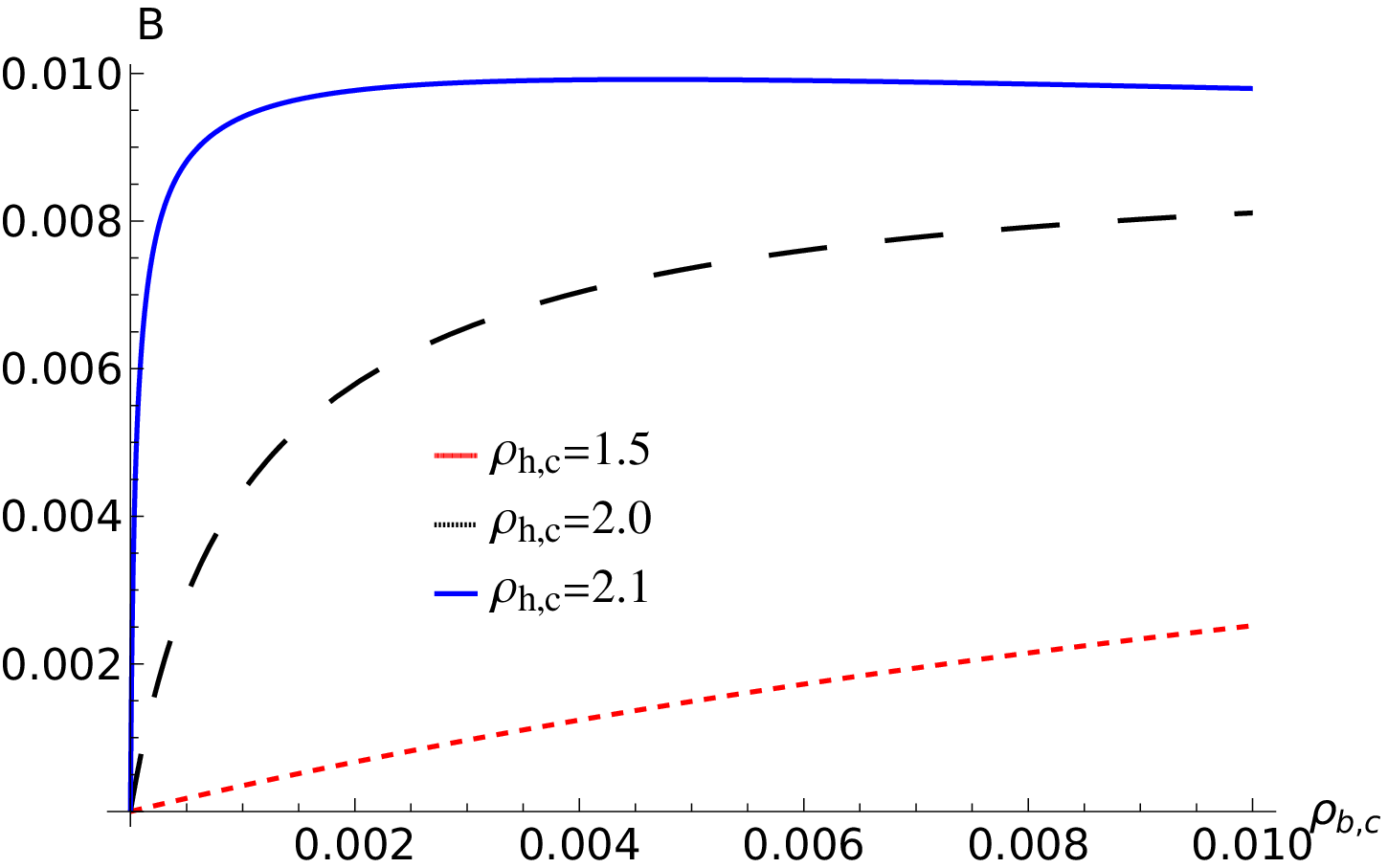}
	\caption{The variation of the dispersive term
$B$, represented by equation $\left(  8b\right)  $ a) against $\vartheta$ for
different values of $\kappa$ at $\rho_{h,c}=1.5$, $\rho_{b,c}=0.003$,
$\theta_{c}=0.04$ with $\theta_{b}=0.05$, b) against $\theta_{b}$ for
different values of $\theta_{c}$ at $\rho_{h,c}=1.5$, $\rho_{b,c}=0.003$,
$\kappa=3$ with $\vartheta=5.0$, c) against $\rho_{b,c}$ for different values
of $\rho_{h,c}$ at $\kappa=3.0$, $\vartheta=5.0$, $\theta_{c}=0.03$ with
$\theta_{b}=0.08$.}
\label{Figure3}
\end{figure}
\begin{figure}[!tbp]
  \centering
  \begin{minipage}[b]{0.45\textwidth}
    \includegraphics[width=\textwidth]{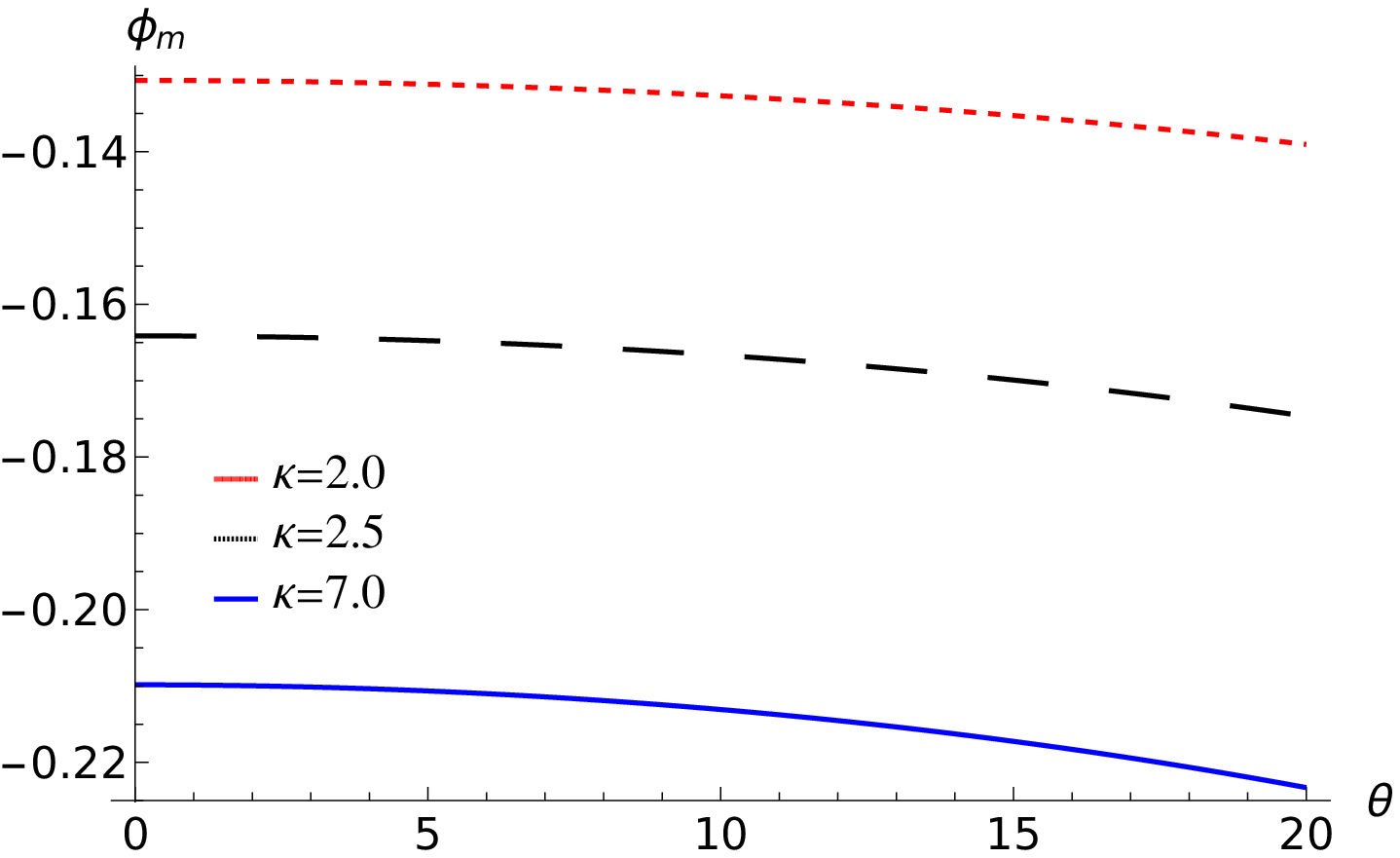}
  \end{minipage}
  \hfill
  \begin{minipage}[b]{0.45\textwidth}
    \includegraphics[width=\textwidth]{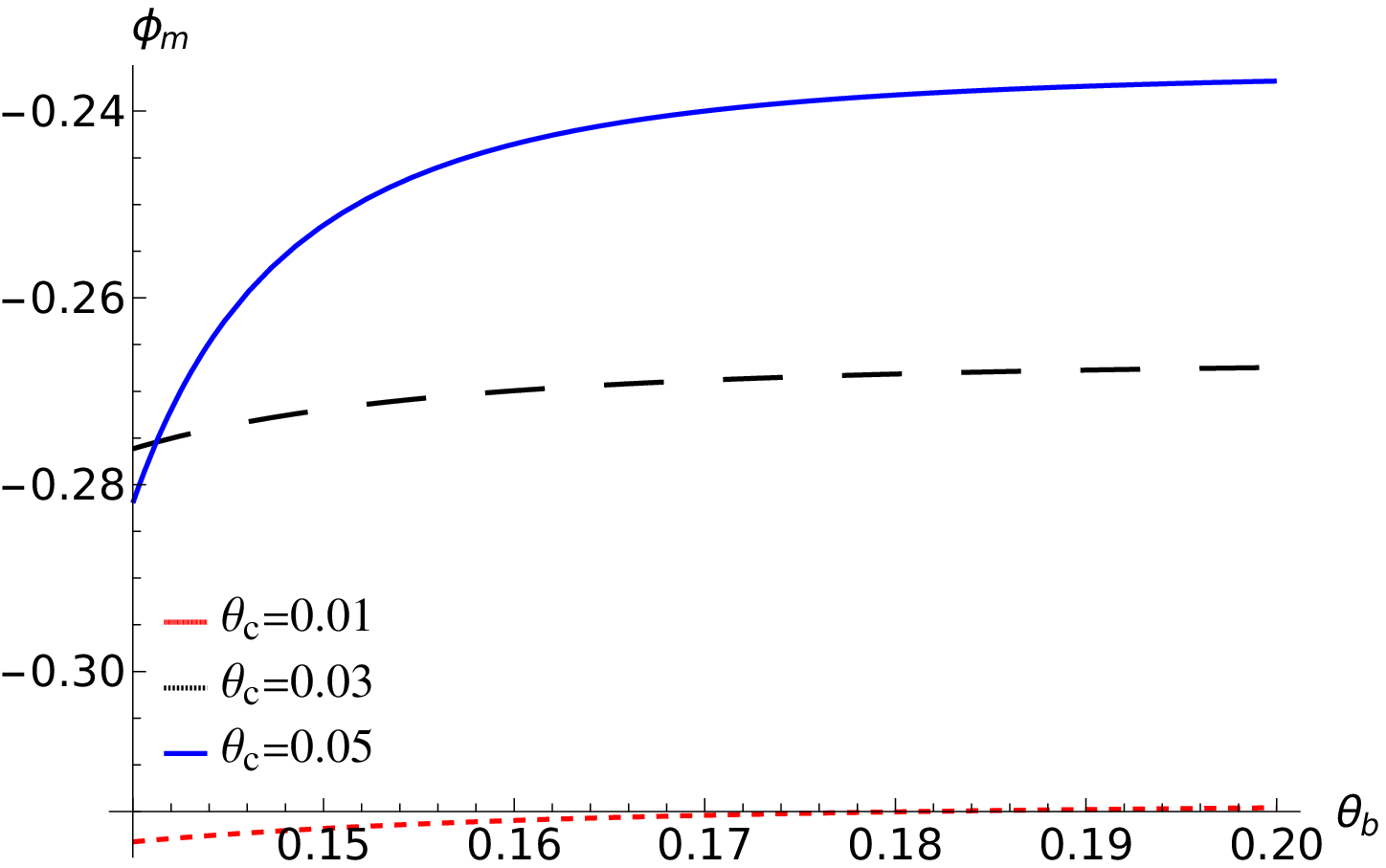}
  \end{minipage}
    \centering
	\includegraphics[width=0.6\linewidth]{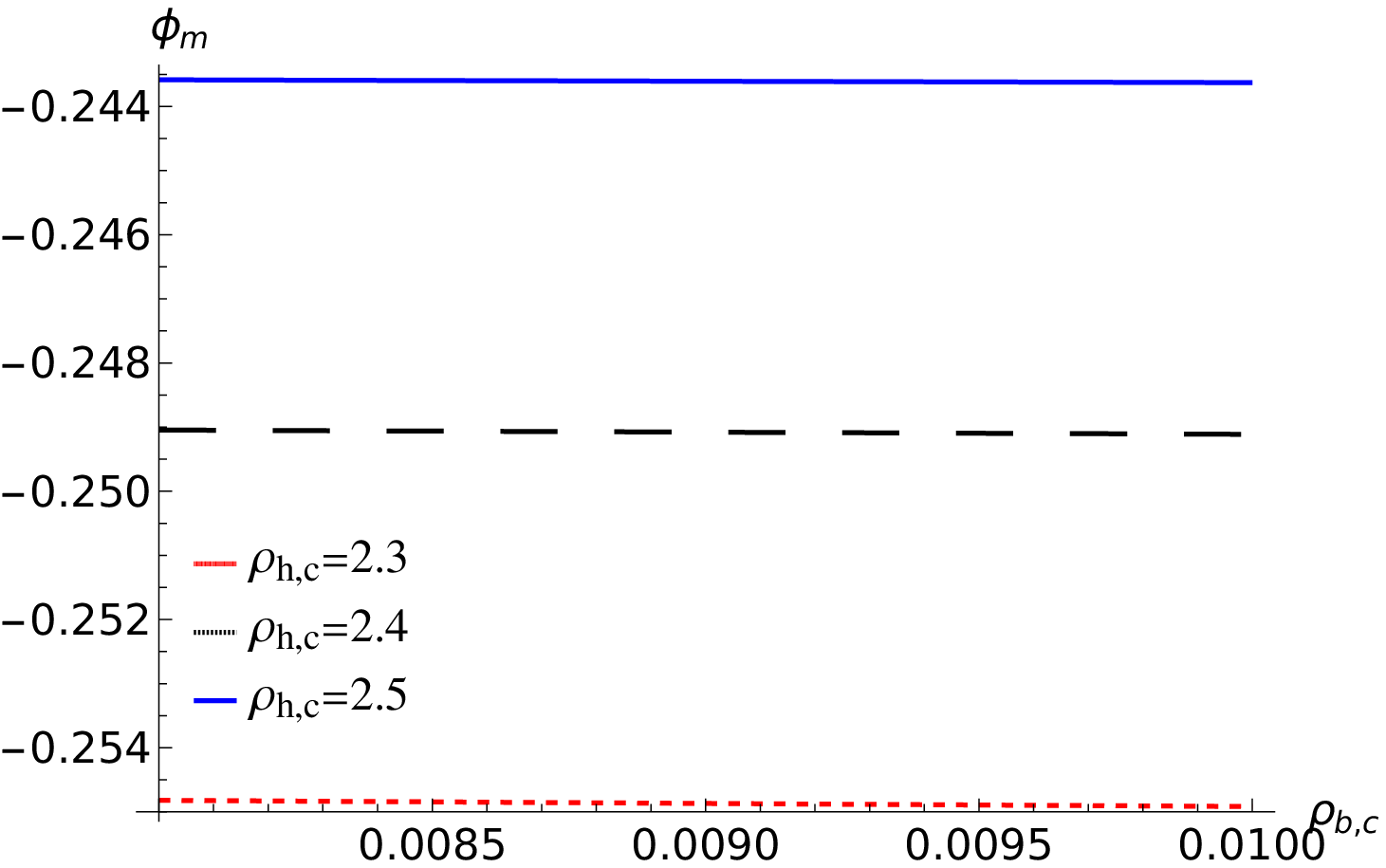}
	\caption{The variation of the EASWs amplitude
$\phi_{m}$, represented by equation $\left(  8b\right)  $ a) against
$\vartheta$ for different values of $\kappa$ at $\rho_{h,c}=2.5$, $\rho
_{b,c}=0.004$, $\theta_{c}=0.04$ with $\theta_{b}=0.15$, b) against
$\theta_{b}$ for different values of $\theta_{c}$ at $\rho_{h,c}=1.5$,
$\rho_{b,c}=0.006$, $\kappa=3$ with $\vartheta=5.0$, c) against $\rho_{b,c}$
for different values of $\rho_{h,c}$ at $\kappa=3.0$, $\vartheta=5.0$,
$\theta_{c}=0.01$ with $\theta_{b}=0.14$.}
\label{Figure4}
\end{figure}
\begin{figure}[!tbp]
  \centering
  \begin{minipage}[b]{0.45\textwidth}
    \includegraphics[width=\textwidth]{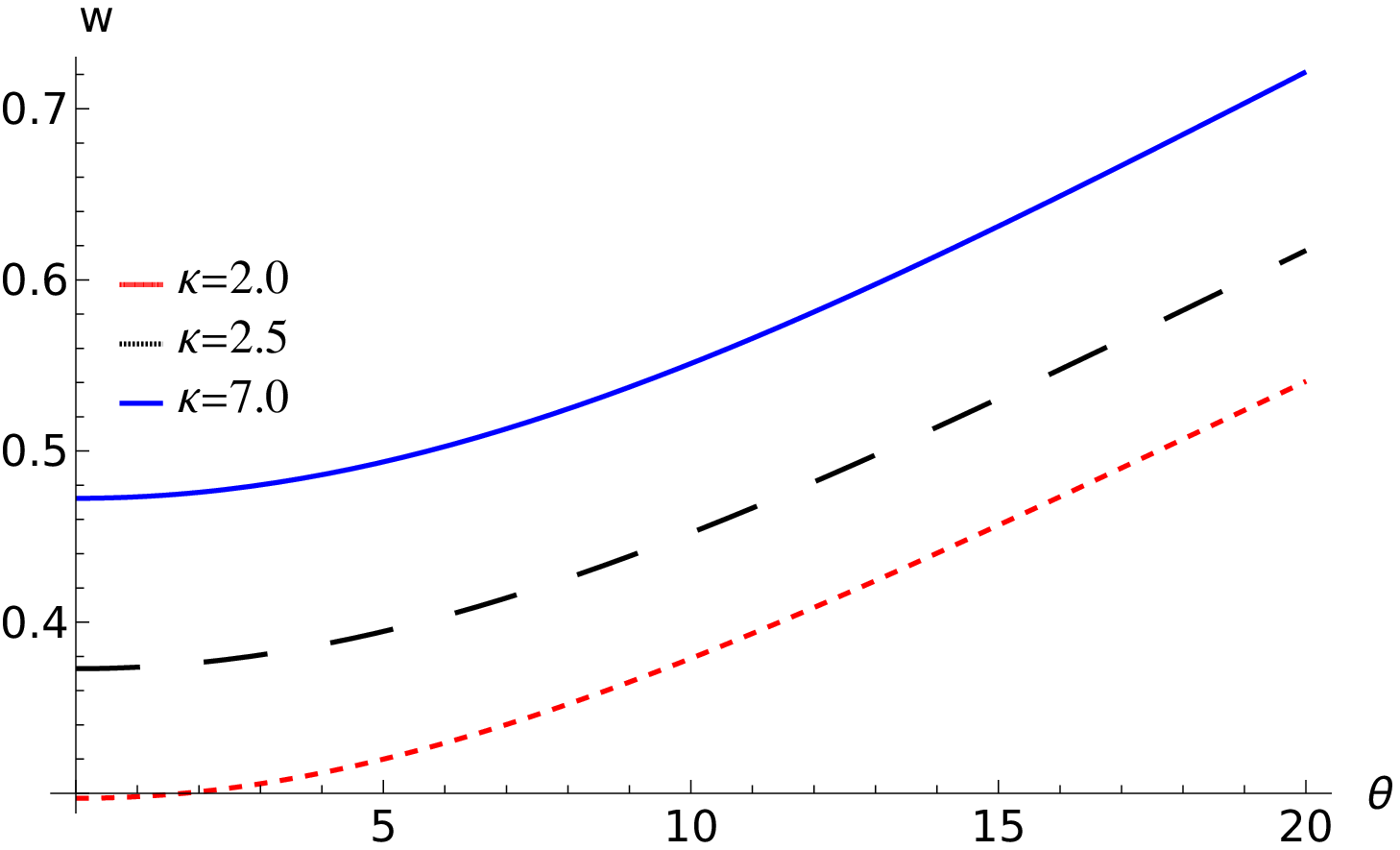}
  \end{minipage}
  \hfill
  \begin{minipage}[b]{0.45\textwidth}
    \includegraphics[width=\textwidth]{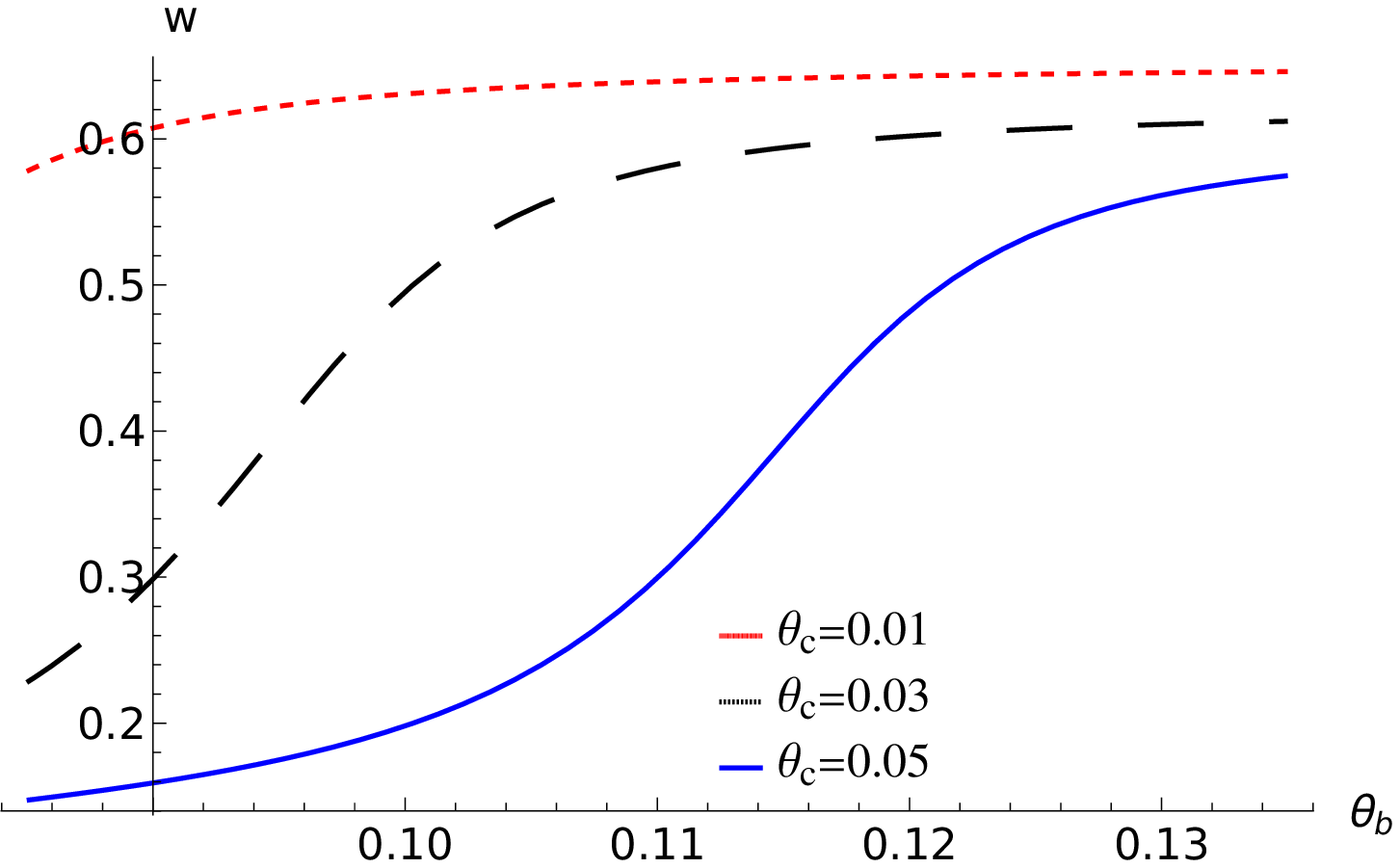}
  \end{minipage}
    \centering
	\includegraphics[width=0.6\linewidth]{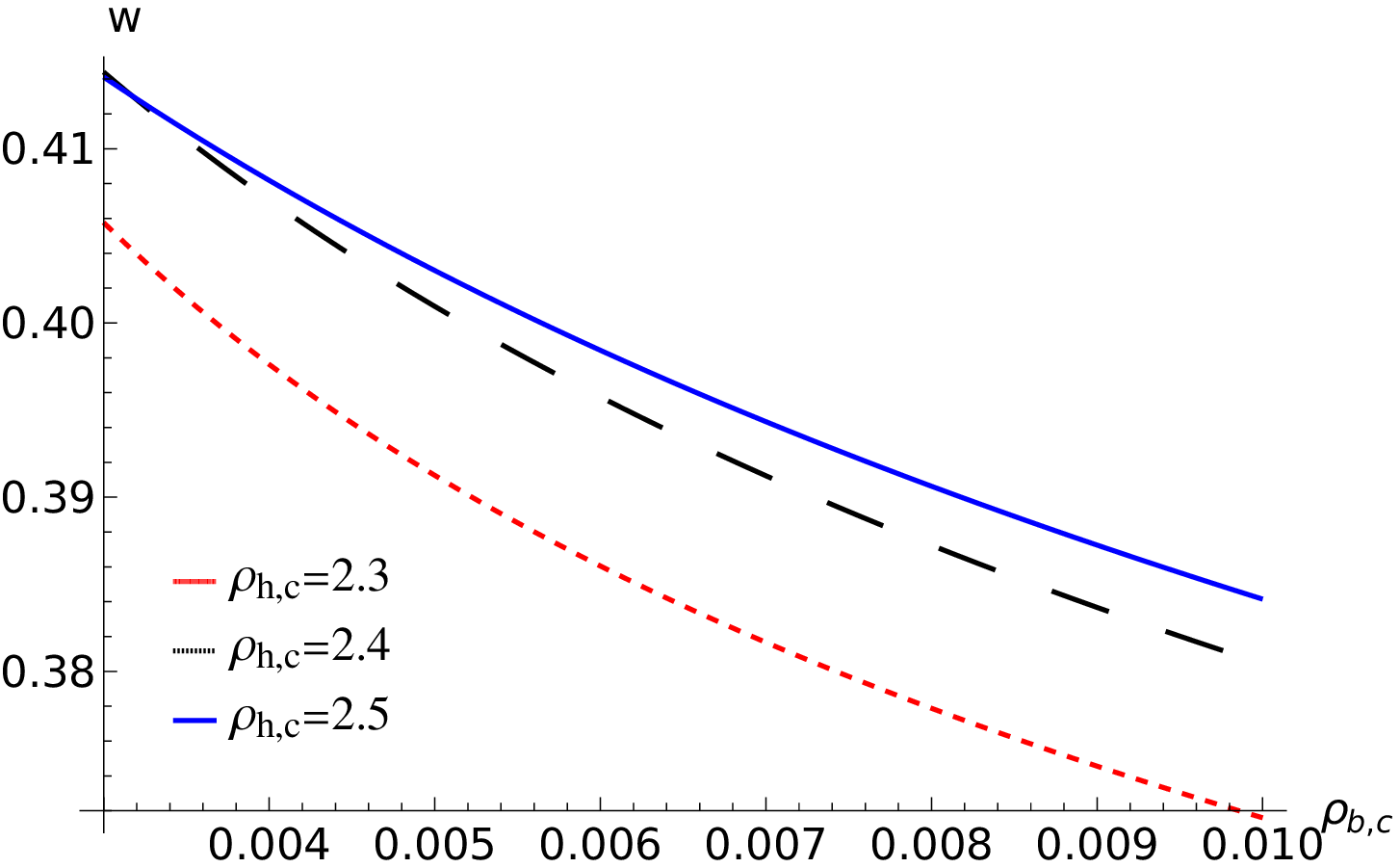}
	\caption{The variation of the EASWs width $w$,
represented by equation $\left(  8b\right)  $ a) against $\vartheta$ for
different values of $\kappa$ at $\rho_{h,c}=2.5$, $\rho_{b,c}=0.004$,
$\theta_{c}=0.04$ with $\theta_{b}=0.15$, b) against $\theta_{b}$ for
different values of $\theta_{c}$ at $\rho_{h,c}=1.5$, $\rho_{b,c}=0.006$,
$\kappa=3$ with $\vartheta=5.0$, c) against $\rho_{b,c}$ for different values
of $\rho_{h,c}$ at $\kappa=3.0$, $\vartheta=5.0$, $\theta_{c}=0.03$ with
$\theta_{b}=0.08$.}
\label{Figure5}
\end{figure}
\begin{figure}[h]
\centering
  \begin{minipage}[b]{0.45\textwidth}
    \includegraphics[width=\textwidth]{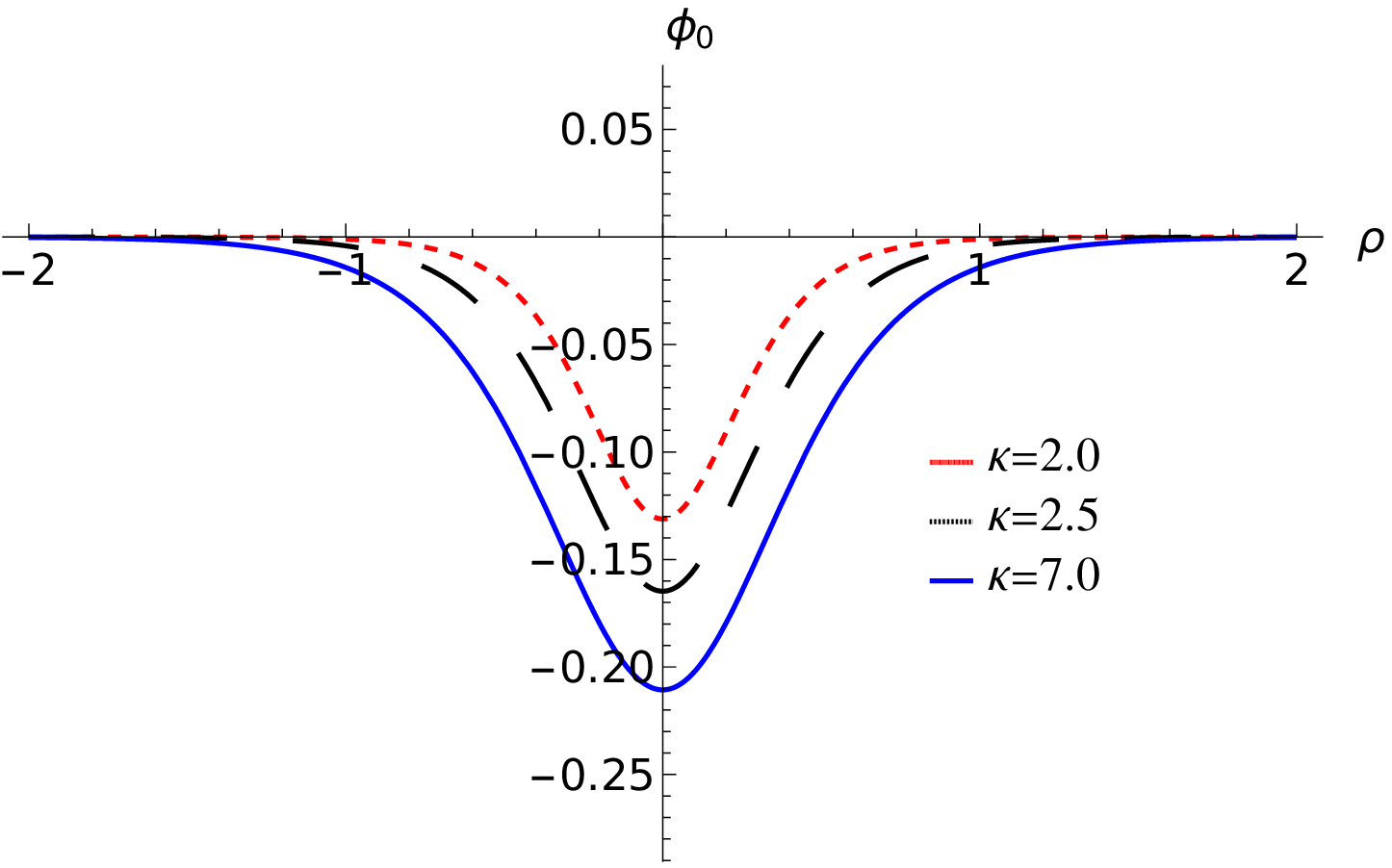}
  \end{minipage}
  \hfill
  \begin{minipage}[b]{0.45\textwidth}
    \includegraphics[width=\textwidth]{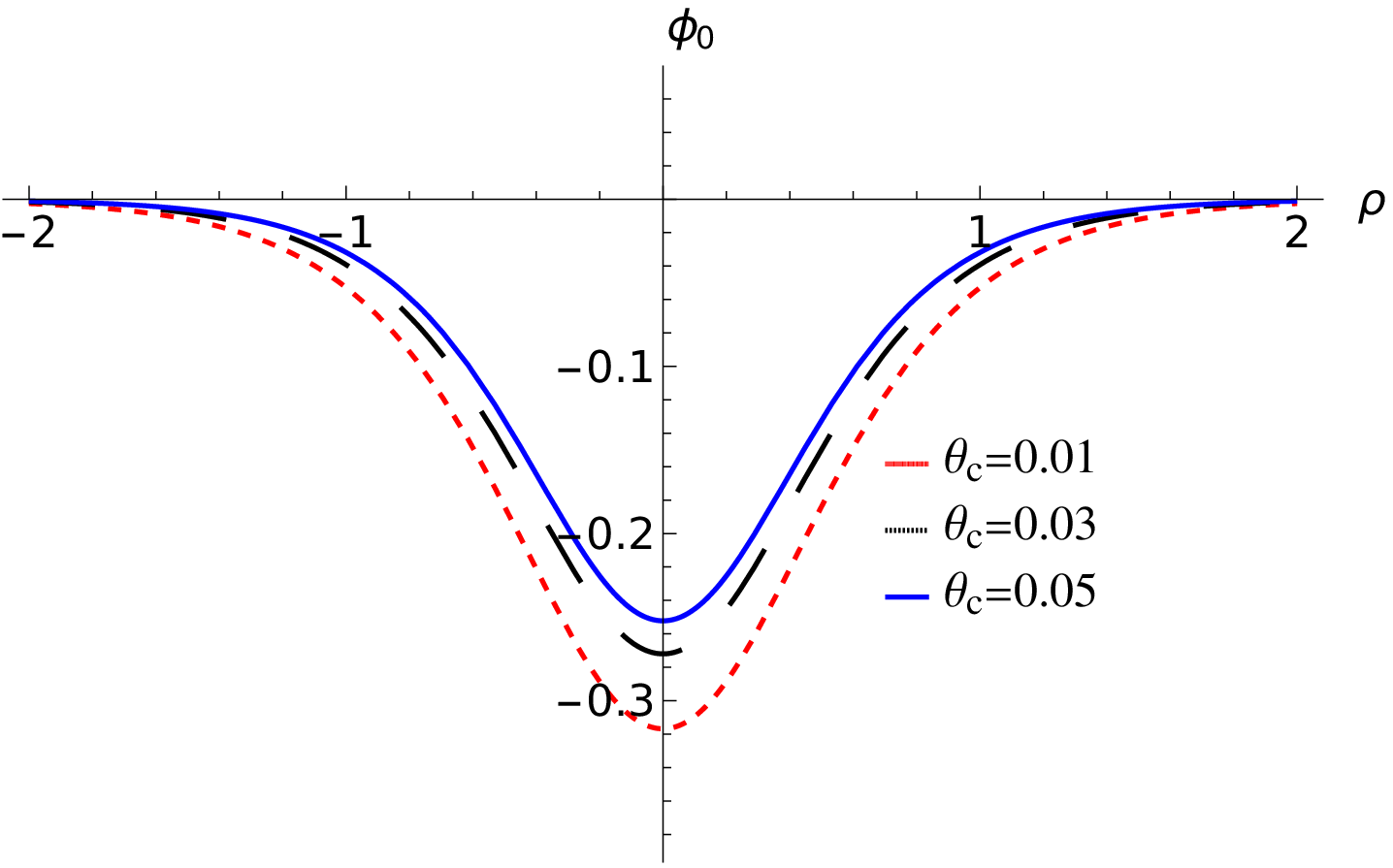}
  \end{minipage}
    \centering
	\begin{minipage}[b]{0.45\textwidth}
    \includegraphics[width=\textwidth]{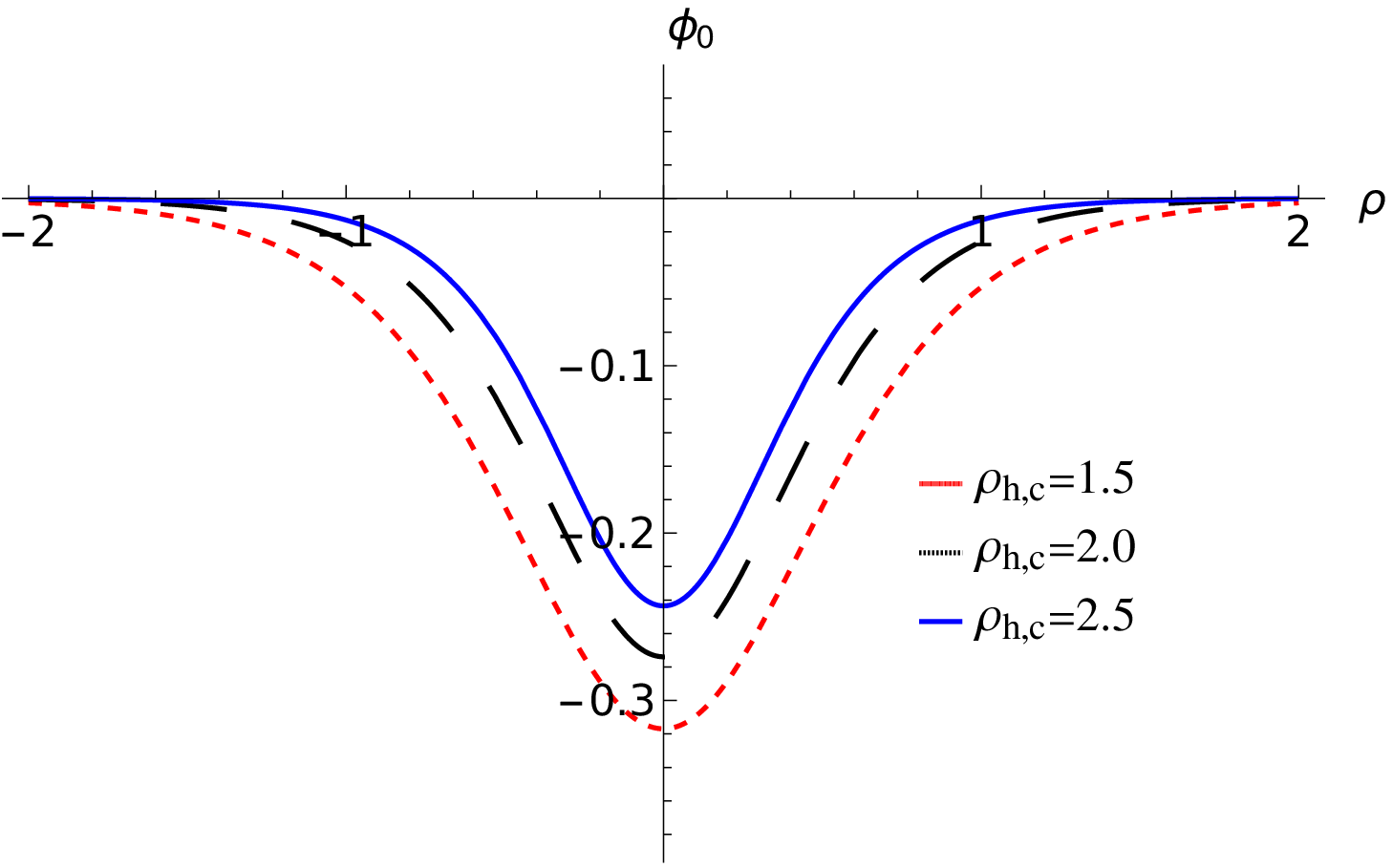}
  \end{minipage}
  \hfill
  \begin{minipage}[b]{0.45\textwidth}
    \includegraphics[width=\textwidth]{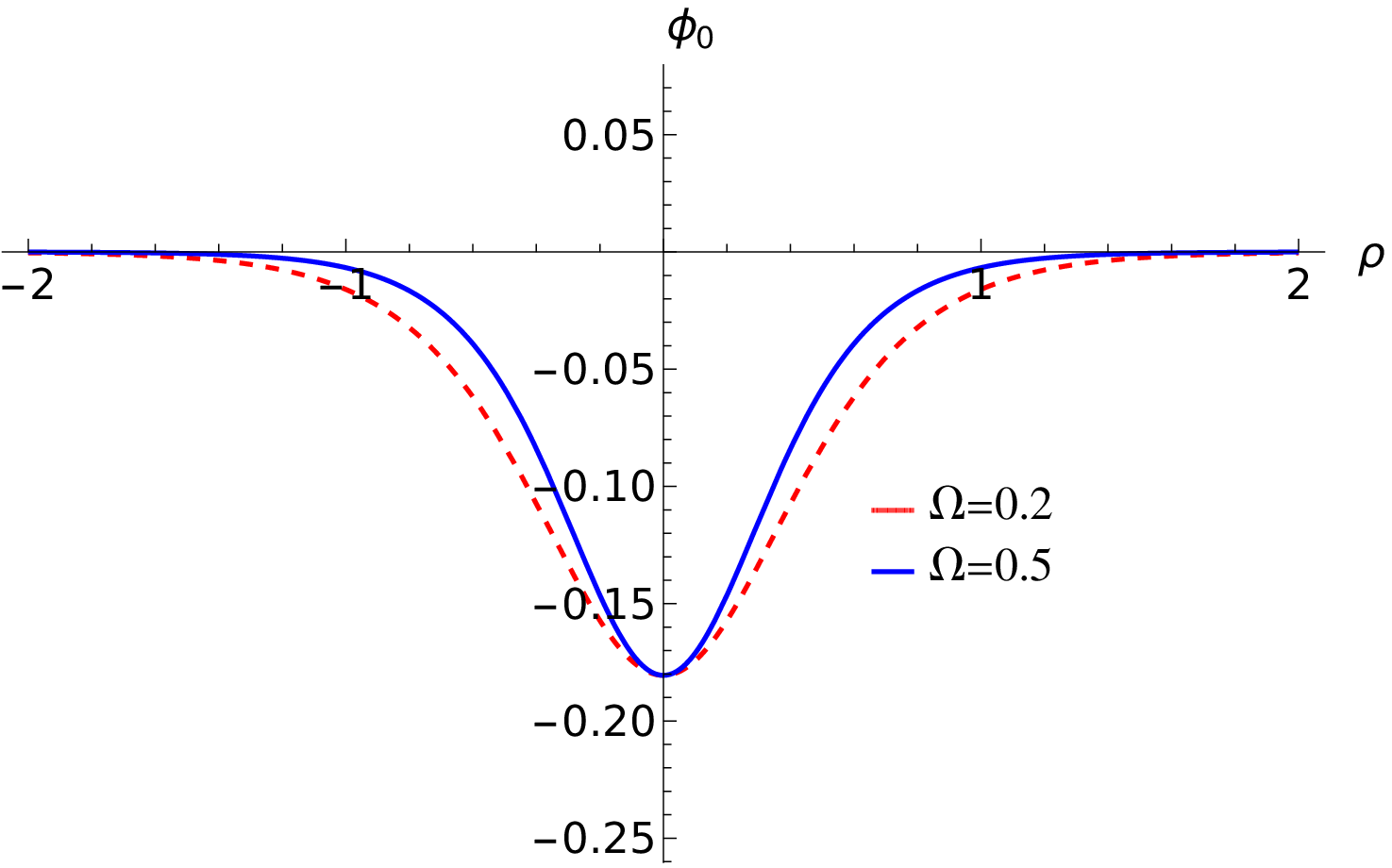}
  \end{minipage}
	\caption{The evolution of $\phi_{0}$ of EASWs
that represented by equation $\left(  13\right)  $ with $\rho$ at $\vartheta=5$
for different values of a) $\kappa$ with $\rho_{h,c}=2.5$, $\Omega=0.5$,
$\rho_{b,c}=0.004$, $\theta_{b}=0.15$, and $\theta_{c}=0.04$, b) $\theta_{c}$
with $\Omega=0.5$, $\rho_{h,c}=1.5$, $\rho_{b,c}=0.006$, $\theta_{b}=0.15$,
and $\kappa=3.0$, c) $\rho_{h,c}$ with $\kappa=3.0$, $\Omega=0.5$, $\rho
_{b,c}=0.004$, $\theta_{b}=0.14$, and $\theta_{c}=0.01$, d) $\Omega$ with
$\rho_{h,c}=2.5$, $\rho_{b,c}=0.004$, $\theta_{b}=0.15$, $\theta_{c}=0.04$,
and $\kappa=3.0$.}
\label{Figure6}
\end{figure}
\begin{figure}[h]
\centering
  \begin{minipage}[b]{0.45\textwidth}
    \includegraphics[width=\textwidth]{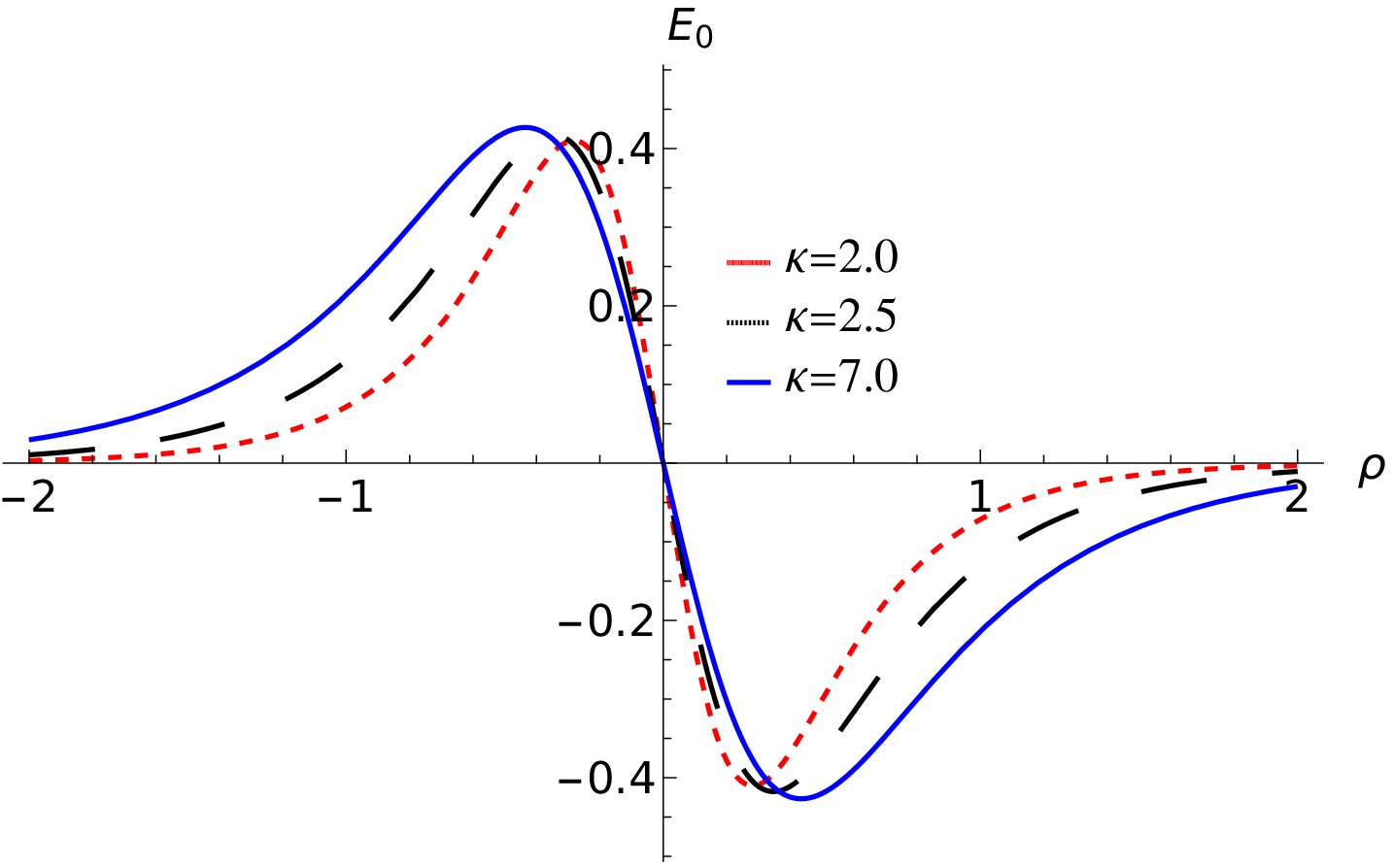}
  \end{minipage}
  \hfill
  \begin{minipage}[b]{0.45\textwidth}
    \includegraphics[width=\textwidth]{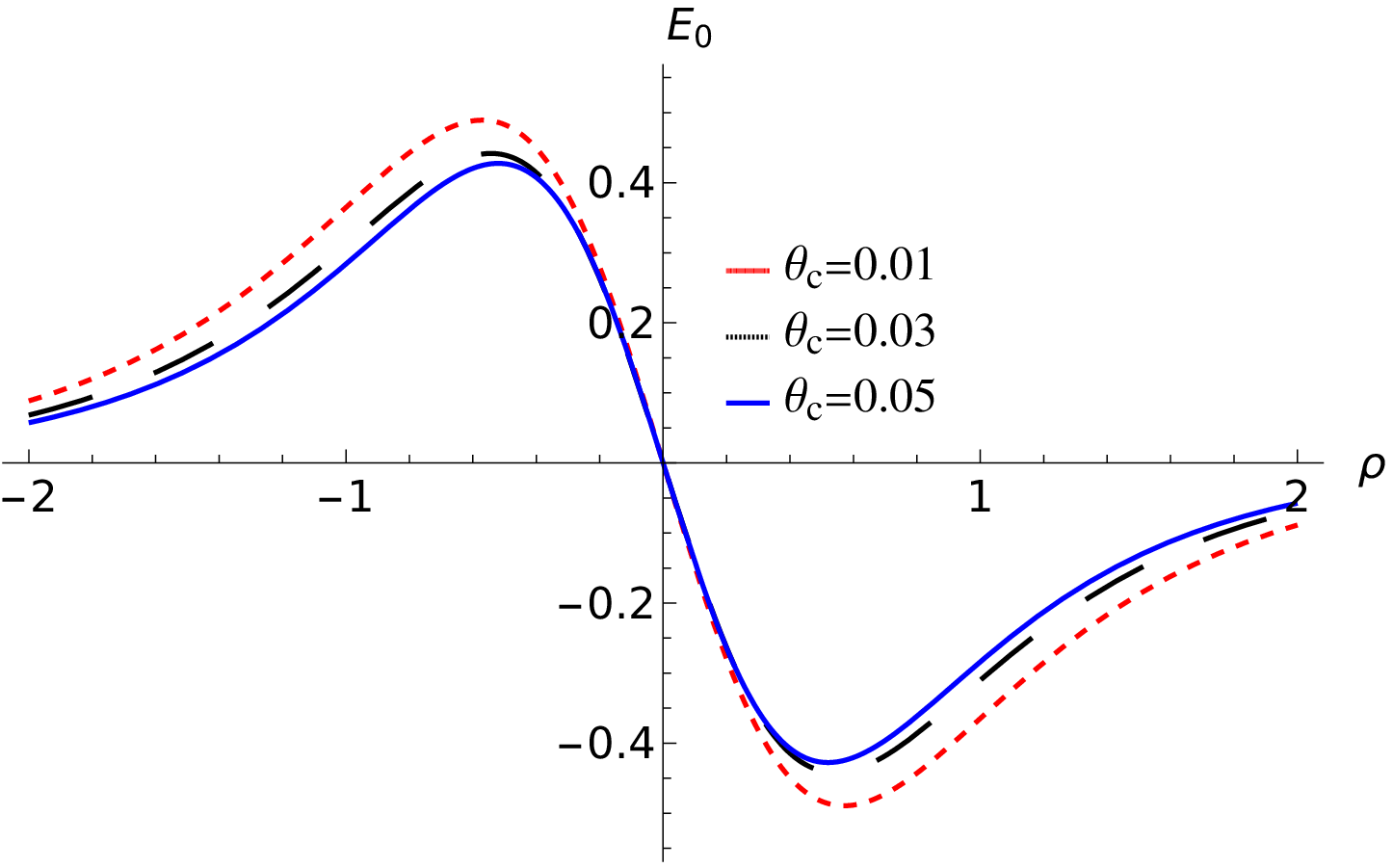}
  \end{minipage}
    \centering
	\begin{minipage}[b]{0.45\textwidth}
    \includegraphics[width=\textwidth]{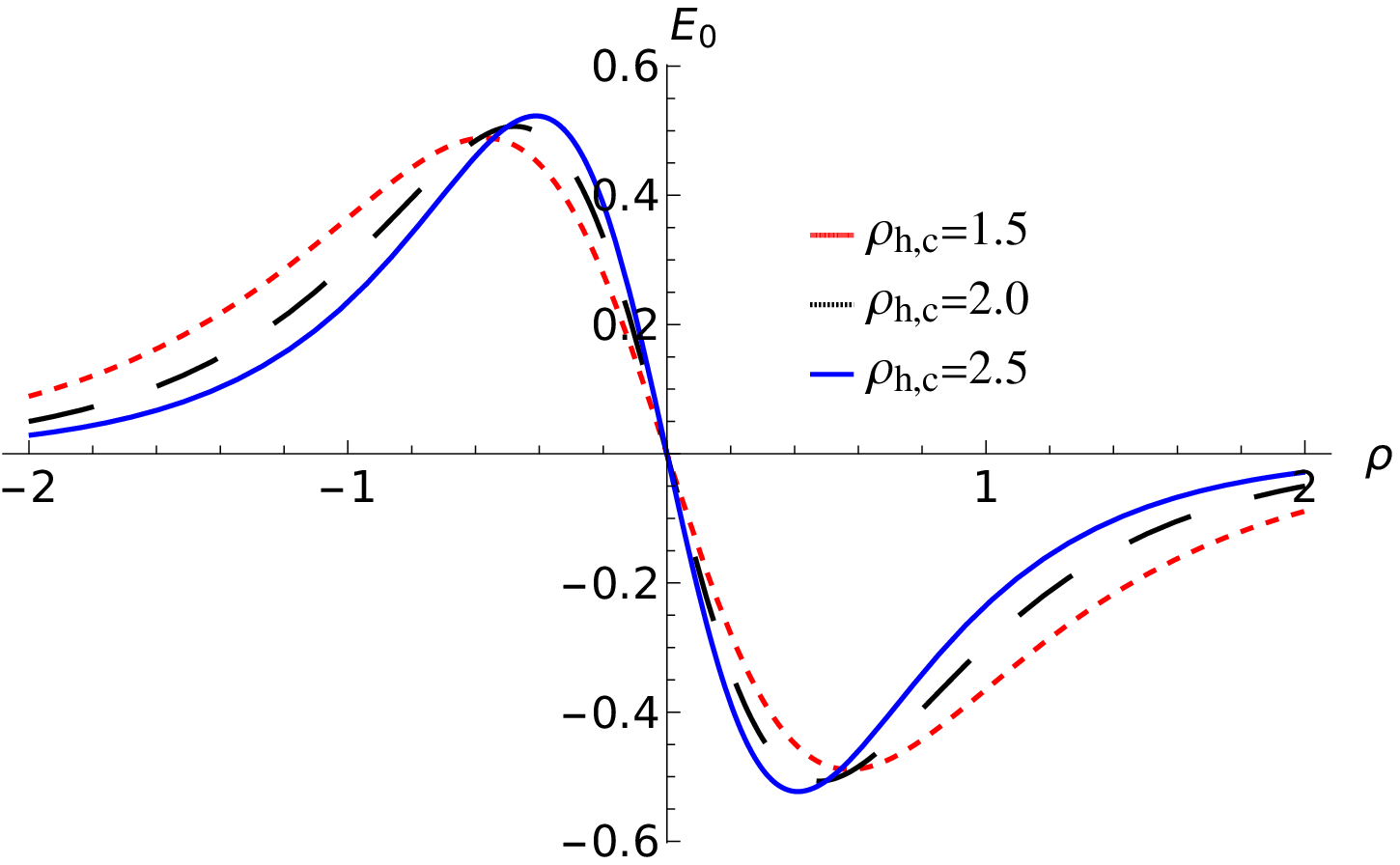}
  \end{minipage}
  \hfill
  \begin{minipage}[b]{0.45\textwidth}
    \includegraphics[width=\textwidth]{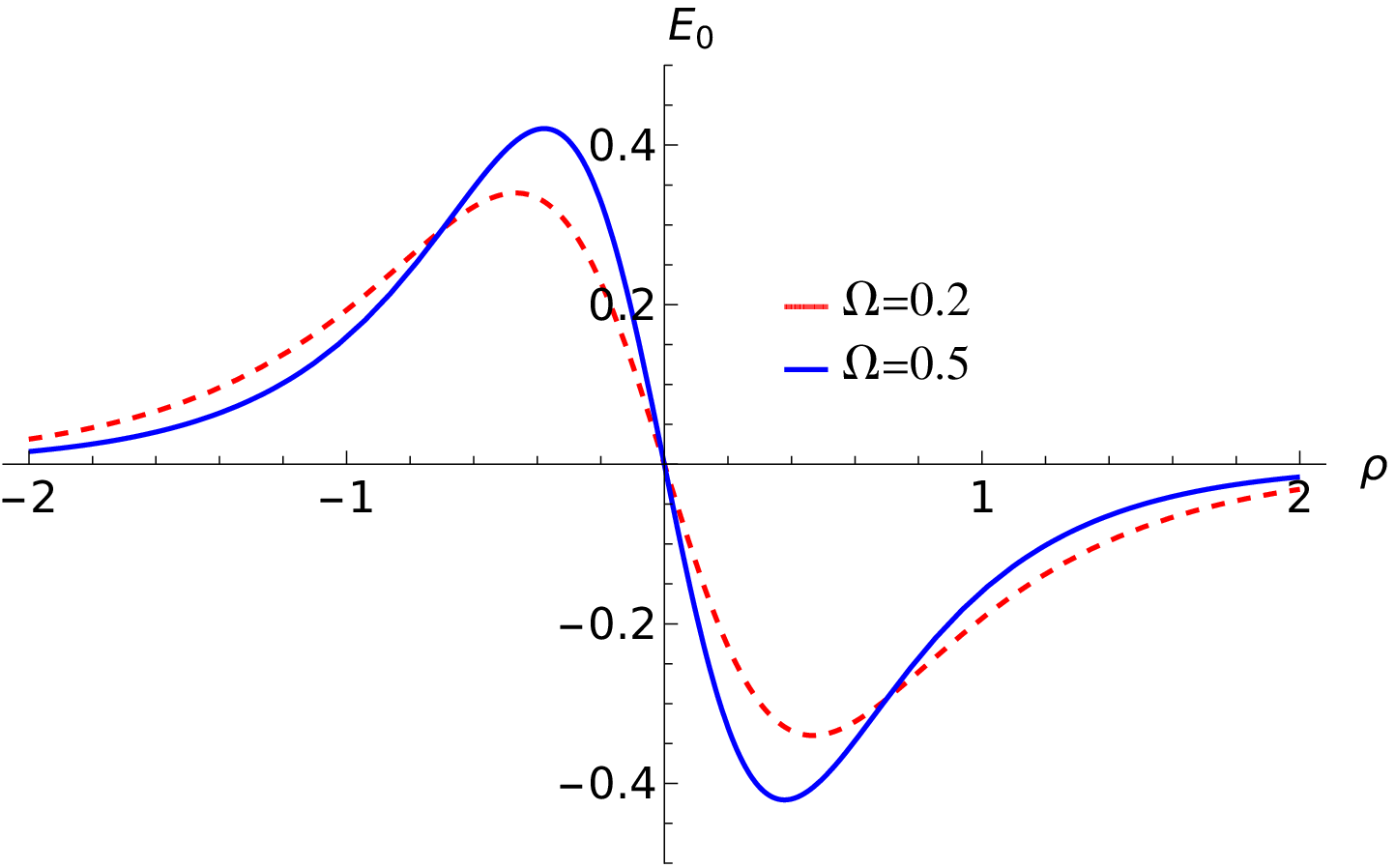}
  \end{minipage}
	\caption{The evolution of the associated
electric field, $E_{0}$ of EASWs that represented by equation $\left(15\right)  $ with $\rho$ for the potentials those represented by Fig. $6$.}
\label{Figure7}
\end{figure}
\begin{figure}[!tbp]
  \centering
  \begin{minipage}[b]{0.45\textwidth}
    \includegraphics[width=\textwidth]{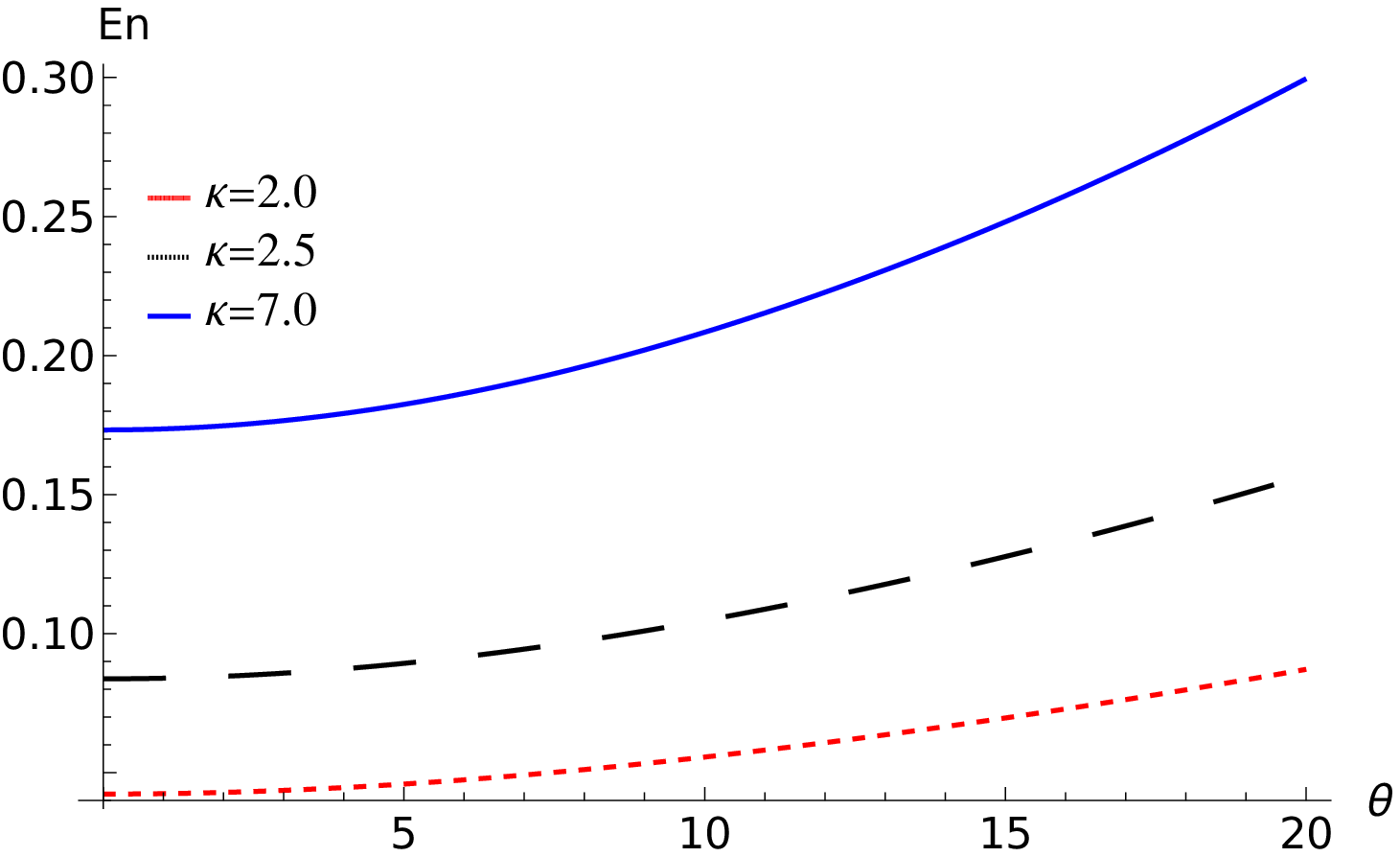}
  \end{minipage}
  \hfill
  \begin{minipage}[b]{0.45\textwidth}
    \includegraphics[width=\textwidth]{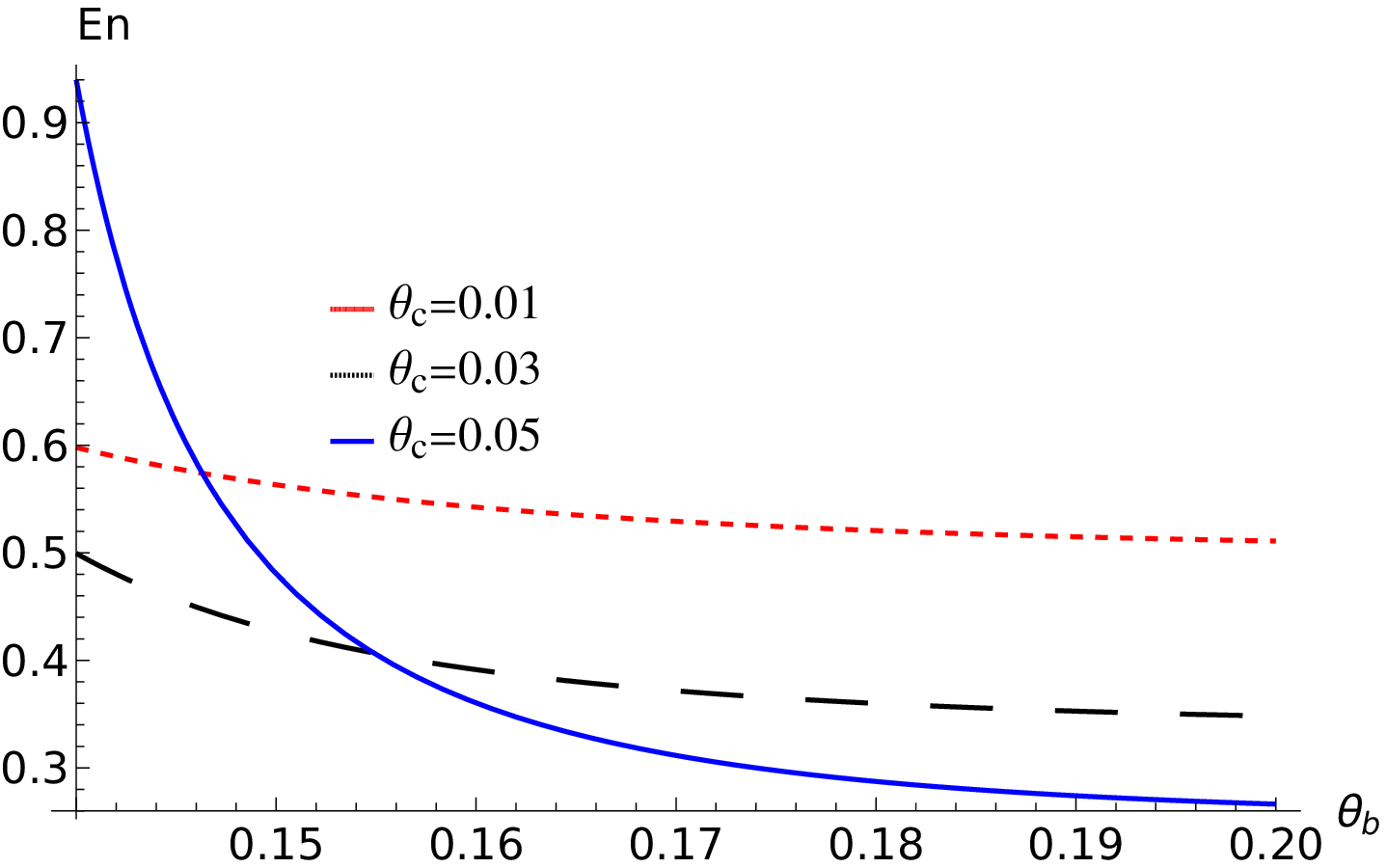}
  \end{minipage}
    \centering
	\includegraphics[width=0.6\linewidth]{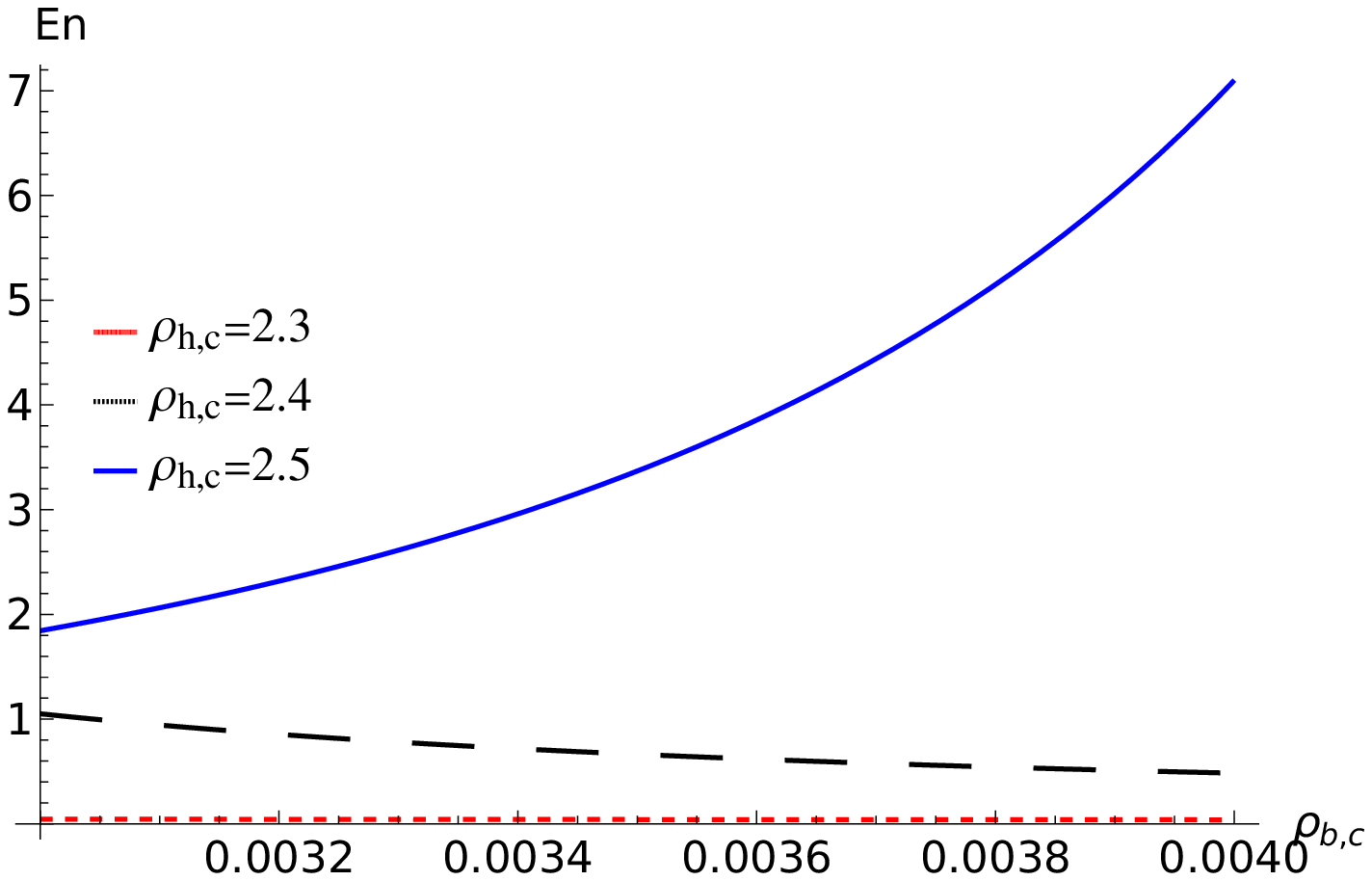}
	\caption{The evolution of the energy $E_{n}$
of the EASWs that represented by equation $\left(  16\right)  $ at
$\Omega=0.5$, and $M$ $=0.4$, a) against $\theta$ for different values of
$\kappa$ at $\rho_{h,c}=2.5$, $\rho_{b,c}=0.003$, $\theta_{c}=0.04$ with
$\theta_{b}=0.15$, b) against $\theta_{b}$ for different values of $\theta
_{c}$ at $\rho_{h,c}=1.5$, $\rho_{b,c}=0.1$, $\kappa=3$ with $\theta=5.0$, c)
against $\rho_{b,c}$ for different values of $\rho_{h,c}$ at $\kappa=3.0$,
$\theta=5.0$, $\theta_{c}=0.03$ with $\theta_{b}=0.08$.}
\label{Figure8}
\end{figure}
\begin{figure}[!tbp]
  \centering
  \begin{minipage}[b]{0.45\textwidth}
    \includegraphics[width=\textwidth]{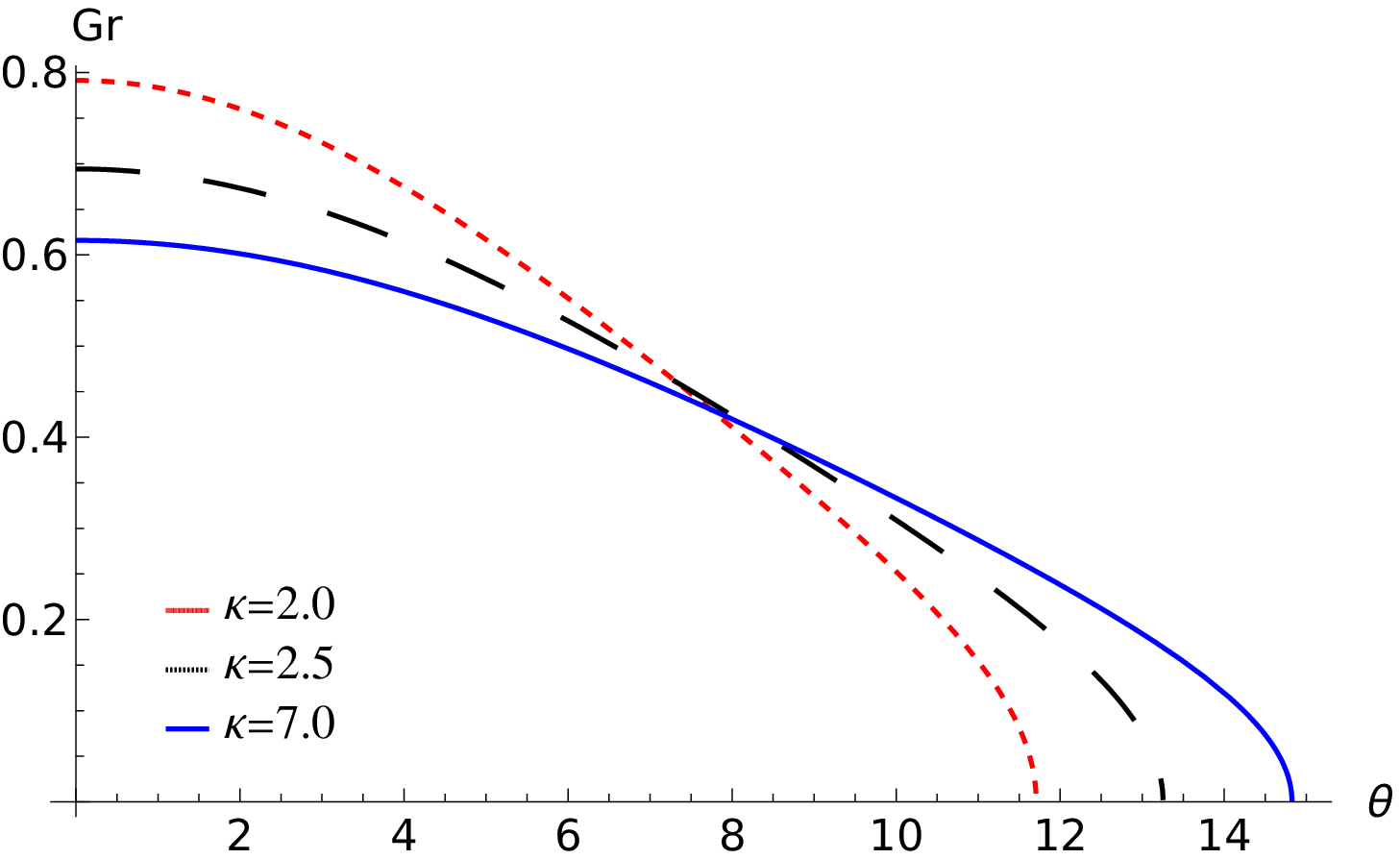}
  \end{minipage}
  \hfill
  \begin{minipage}[b]{0.45\textwidth}
    \includegraphics[width=\textwidth]{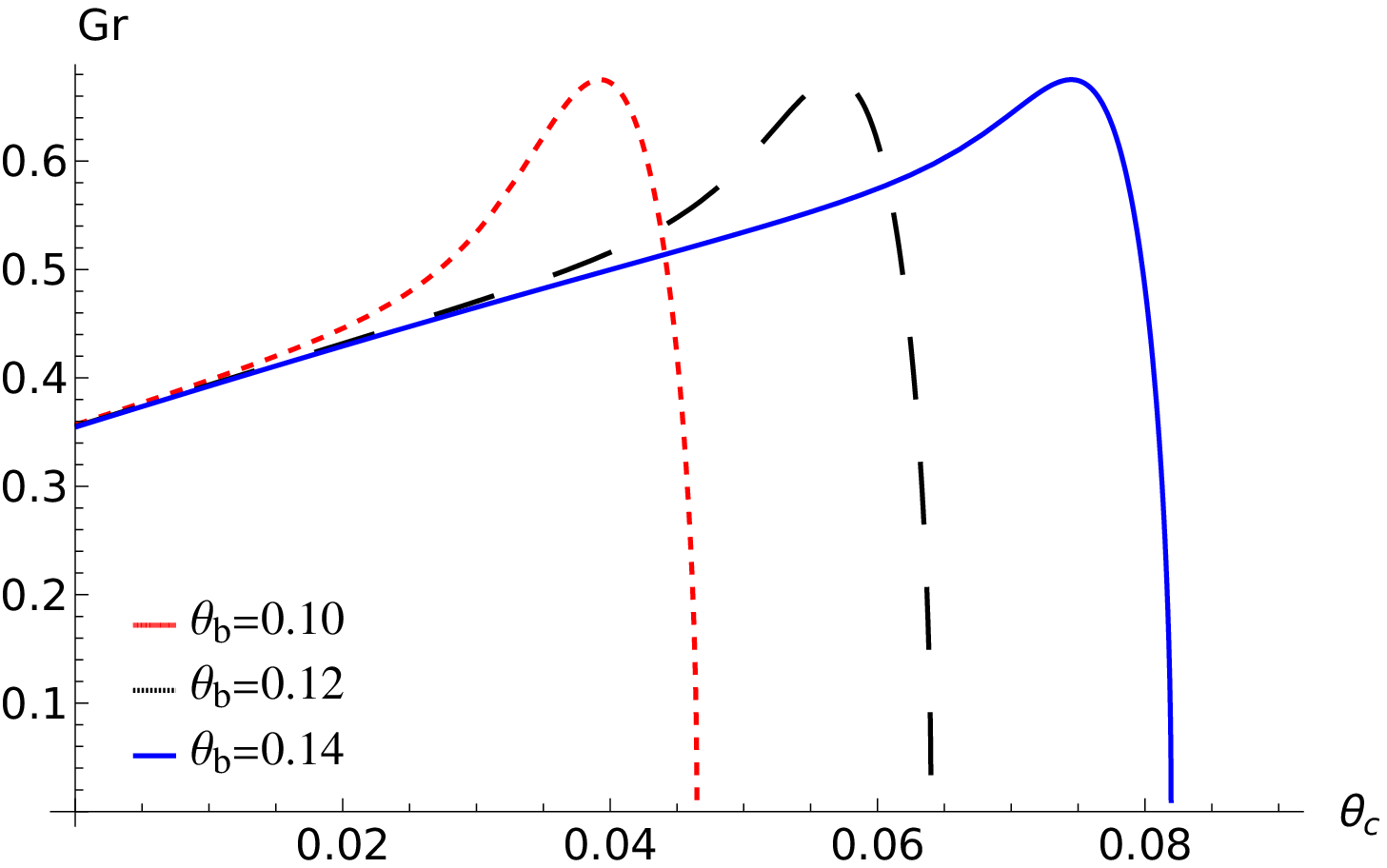}
  \end{minipage}
    \centering
	\includegraphics[width=0.6\linewidth]{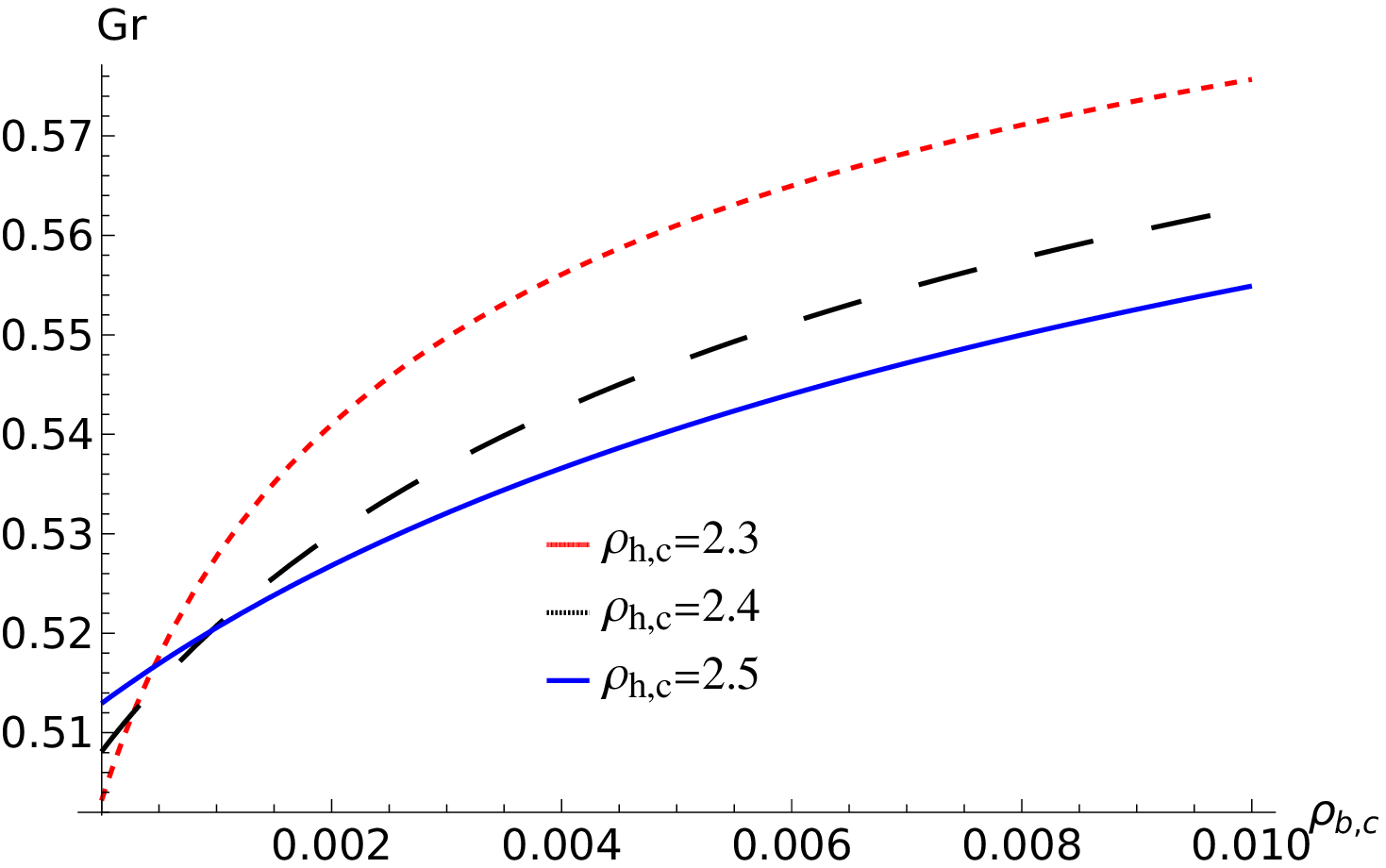}
	\caption{The variation of the growth rate,
$Gr$, that represented by equation $\left(  27\right)  $ at $l_{\xi
}=0.7,l_{\eta}=0.4$, $\Omega=0.5$, and $M$ $=0.4$, a) against $\theta$ for
different values of $\kappa$ at $\rho_{h,c}=2.5$, $\rho_{b,c}=0.004$,
$\theta_{c}=0.04$ with $\theta_{b}=0.15$, b) against $\theta_{b}$ for
different values of $\theta_{c}$ at $\rho_{h,c}=1.5$, $\rho_{b,c}=0.006$,
$\kappa=3$ with $\theta=5.0$, c) against $\rho_{b,c}$ for different values of
$\rho_{h,c}$ at $\kappa=3.0$, $\theta=5.0$, $\theta_{c}=0.03$ with $\theta
_{b}=0.08$.}
\label{Figure9}
\end{figure}


\begin{thebibliography}{99}                                         
\bibitem {rr1} W. Gonzalez, E. Parker, Magnetic reconnection: Concepts and Applications, Springer, London, 2016. 

 \bibitem {r2004} A. Vaivads, Y. Khotyaintsev, M. Andre. et al., Phys. Rev. Lett. 93 (2004) 105001. 
 
\bibitem {r2020} L. C. Lee, K. H. Lee, Rev. Mod. Plasma Phys. 4 (2020) 9.

\bibitem{Froment2021}C. Froment, V. Krasnoselskikh, T. Dudok de Wit et al., A\&A 650 (2021) A5.

\bibitem{Lavraud2021}B. Lavraud, R. Kieokaew, N. Fargette et al., A\&A 656 (2021) A37. 

\bibitem{Goetz1991}J. A. Goetz, R. N. Dexter, S. C. Prager, Phys. Rev. Lett. 66 (1991) 608-611. 

\bibitem {r1}D. A. Gurnett et al., Sci. 262 (1993) 199–203.

\bibitem {r2} M. Hamrin et al., Ann Geophys 20 (2002) 1943–1958.

\bibitem {r3}D. A. Gurnett et al., Geophys. Res. Lett. 30 (2003) 2209. 

\bibitem {Farrell2002}W. M. Farrell, M. D. Desch, M. L. Kaiser, K. Goetz, Geophys. Res. Lett. 29 (2002) 1902.

\bibitem {Deng2004} X. H. Deng, H. Matsumoto, H. Kojima, T. Mukai, R. R. Anderson, W. Baumjohann, R. Nakamura, J. Geophys. Res. 109 (2004) A05206.

\bibitem{Viberg2013} H. Viberg, Yu. V. Khotyaintsev, A. Vaivads, M. André, and J. S. Pickett,
Geophys. Res. Lett. 40 (2013) 1032–1037.

\bibitem{Deng2006}X. H. Deng, R. X. Tang a , H. Matsumoto, J. S. Pickett, A. N. Fazakerley, H. Kojima, W. Baumjohann, A. Coates, R. Nakamura, D. A. Gurnett, Z. X. Liu, Adv. Space Res. 37 (2006) 1373–1381.

\bibitem{Wei2007} X. H. Wei, J. B. Cao, G. C. Zhou, O. Santolik, H. Reme, I. Dandouras, N. Cornilleau-Wehrlin, E. Lucek, C. M. Carr, A. Fazakerley, J. Geophys. Res. 112 (2007) A10225.

\bibitem{Burch2016}J. L. Burch, R. B. Torbert, T. D. Phan, L. J. Chen, T. E. Moore, R. E. Ergun et al., Sci. 352 (2016) 6290.

\bibitem{Le Contel2016}O. Le Contel, A. Retinò, H. Breuillard, L. Mirioni, P. Robert, A. Chasapis et al., Geophys. Res. Lett. 43 (2016) 5943–5952. 

\bibitem{Wilder2016}F. D. Wilder, R. E. Ergun, S. J. Schwartz, D. L. Newman, S. Eriksson, J. E. Stawarz et al., Geophys. Res. Lett. 43 (2016) 8859–8866. 

\bibitem{Ergun2016}R. E. Ergun, J. C. Holmes, K. A. Goodrich, F. D. Wilder, J. E. Stawarz, S. Eriksson et al., Geophys. Res. Lett. 43 (2016) 5626–5634. 

\bibitem{Wilder2019}F. D. Wilder, R. E. Ergun, S. Hoilijoki, J. Webster, M. R. Argall, N. Ahmadi et al., J. Geophys. Res. Space Phys. 124 (2019) 7837–7849.

\bibitem{Zhou2016} M. Zhou et al., Geophys. Res. Lett. 43 (2016) 4808–4815.

\bibitem{Khotyaintsev2020} Y. V. Khotyaintsev, D. B. Graham, K. Steinvall, L. Alm, A. Vaivads, A. Johlander, C. Norgren, W. Li, A. Divin, H. S. Fu, K.-J. Hwang, J. L. Burch, N. Ahmadi, O. Le Contel, D. J. Gershman, C. T. Russell, and R. B. Torbert, Phys. Rev. Lett. 124 (2020) 045101.

\bibitem{Drake2003} J. F. Drake, M. Swisdak, C. Cattell, M. A. Shay, B. N. Rogers, A. Zeiler, Sci. 299 (2003) 873.

\bibitem{Yu2021} X. Yu, Q. Lu, R. Wang et al., J. Geophys. Res. Space Phys. 126 (2021).

\bibitem{Wei2007} X. H. Wei, J. B. Cao, G. C. Zhou, O. Santolik, H. Reme, I. Dandouras, N. Cornilleau-Wehrlin, E. Lucek, C. M. Carr, A. Fazakerley, J. Geophys. Res. 112 (2007) A10225.

\bibitem{Graham2022} D. B. Graham, O. Le Contel, K.-J. Hwang, Yu. V. Khotyaintsev 1 , M. André, A. Vaivads, A. Divin, J. F. Drake, C. Norgren, P.-A. Lindqvist, A. C. Rager, D. J. Gershman, C. T. Russell, J. L. Burch, K. Dokgo, Nat. Commun. 13 (2022) 2954.

\bibitem{Cattell2002} C. Cattell, J. Crumley, J. Dombeck, and J. Wygant, J. Geophys. Res. 29 (2002) 1065.

\bibitem{Graham2014} D. B. Graham, Y. V. Khotyaintsev, A. Vaivads, M. André, A. N. Fazakerley, Phys. Rev. Lett. 112 (2014) 215004.

\bibitem{Graham2015} D. B. Graham, Y. V. Khotyaintsev, A. Vaivads, and M. André, Geo- phys. Res. Lett. 42 (2015) 215–224.


\bibitem{Zakharov1974}V. E. Zakharov, E. A. Kuznetsov, Sov. Phys. JEPT 39 (1974) 285.

\bibitem{Danehkar2018} A. Danehkar, Plasma Phys. Control. Fusion 60 (2018) 065010.

\bibitem{Hellberg2009} M. A. Hellberg, R. L. Mace, T. K. Baluku, I. Kourakis, N. S. Saini, Phys. Plasmas 16 (2009) 094701.

\bibitem{Sultana2012} S. Sultana, G. arri, I. Kourakis, plasmas. Phys. Plasmas 19 (2012) 012310.

\bibitem{Atteya2018} A. Atteya, S. Sultana, S. Schlickeiser, Chin. J. Phys. 56 (2018) 1931-1939.

\bibitem{Allen1993} M.A. Allen, G. Rowlands, J. Plasma Phys. 50 (1993) 413.

\bibitem{Allen1995} M.A. Allen, G. Rowlands, J. Plasma Phys. 53 (1995) 63.

\bibitem{Mamun1998} A.A. Mamun, Phys. Scr. 58 (1998) 505.

\bibitem{Haider2012} M.M. Haider, A.A. Mamun, Phys. Plasmas 19 (2012) 102105.

\bibitem{Afify2021} M. S. Afify, I. S. Elkamash, M. Shihab, W. M. Moslem, Adv. Space Res. 67 (2021) 4110-4120.

\bibitem{El-Monier2021} S. Y. El-Monier, A. Atteya, Waves Random Complex Media (2021).

\bibitem{Mamun1998} A.A. Mamun, Phys. Scr. 58 (1998) 505.

\bibitem{El-Labany2013} S.K. El-Labany, W. F. El-Taibany, E. E. Behery, Phys. Rev. E 88 (2013) 023108.

\bibitem{Graham2016} D. B. Graham, A. Vaivads, Y. V. Khotyaintsev, and M. André, J. Geophys. Res. Space Physics 121 (2016) 1934–1954.

\bibitem{Fortov2005}  V.E. Fortov, A.V. Ivlev, S.A. Khrapak, G.E. Morfill,
Phys. Rep. 421 (2005) 1.

\bibitem{Zedan2020} N.A. Zedan, A. Atteya, W.F. El-Taibany, S.K. EL-Labany,
Waves Random and Complex Media (2020).

\bibitem{PARIS1973} R. B. Paris, Plasma Physics 15 (1973) 853-870. 

\bibitem{Shukla2002} P. K. Shukla, Phys. Plasmas 9 (2002) 4082.

\bibitem{Shi2020} C. Shi, M. Velli, F. Pucci, A. Tenerani, M. E. Innocenti, ApJ 902 (2020) 142.

\bibitem{Mayur2022}M. R. Bakrania, I. J. Rae, A. P. Walsh, D. Verscharen, A. W. Smith, C. Forsyth, A. Tenerani, Front. Astron. Space Sci. (2022). 

\bibitem{El-Labany2020} S. K. El-Labany, W. M. Moslem, N. K. Elneely, Adv. Space Res. 66 (2020) 266.


\end{thebibliography}
\end{document}